\documentclass{jfm}
\usepackage{graphicx}
\usepackage{epstopdf, epsfig,subfloat,subfig}

\usepackage{amsfonts}
\usepackage{amsmath}
\usepackage{amssymb,color}

\shorttitle{Apparent permeability of porous media}
\shortauthor{Wu, Ho, Germanou, Gu, Liu, Xu, and Zhang }

\title{On the apparent permeability of porous media in rarefied gas flows}

\author{Lei Wu\aff{1}
	\corresp{\email{lei.wu.100@strath.ac.uk}},
	Minh Tuan Ho\aff{1},
	Lefki Germanou\aff{1},
	Xiao-Jun Gu\aff{2},
	Chang Liu\aff{3},
	Kun Xu\aff{3},
	\and Yonghao Zhang\aff{1}
	}

\affiliation{
	\aff{1}James Weir Fluids Laboratory, Department of Mechanical and Aerospace Engineering, University of Strathclyde, Glasgow G1 1XJ, UK
	\aff{2}Computational Science and Engineering Department, STFC Daresbury Laboratory, Warrington WA4 4AD, UK
	\aff{3}Department of Mathematics, Hong Kong University of Science and Technology, Clear Water Bay, Hong Kong, China
}

\begin{document}

\maketitle

\begin{abstract}

The apparent gas permeability of the porous medium is an important parameter in the prediction of unconventional gas production, which was first investigated systematically by Klinkenberg in 1941 and found to increase with the reciprocal mean gas pressure (or equivalently, the Knudsen number). Although the underlying rarefaction effects are well-known, the reason that the correction factor in Klinkenberg's famous equation decreases when the Knudsen number increases has not been fully understood. Most of the studies idealize the porous medium as a bundle of straight cylindrical tubes, however, according to the gas kinetic theory, this only results in an increase of the correction factor with the Knudsen number, which clearly contradicts Klinkenberg's experimental observations. Here, by solving the Bhatnagar-Gross-Krook equation in simplified (but not simple) porous media, we identify, for the first time, two key factors that can explain Klinkenberg's experimental results: the tortuous flow path and the non-unitary tangential momentum accommodation coefficient for the gas-surface interaction. Moreover, we find that Klinkenberg's results can only be observed when the ratio between the apparent and intrinsic permeabilities is $\lesssim30$; at large ratios (or Knudsen numbers) the correction factor increases with the Knudsen number. Our numerical results could also serve as benchmarking cases to assess the accuracy of macroscopic models and/or numerical schemes for the modeling/simulation of rarefied gas flows in complex geometries over a wide range of gas rarefaction. Specifically, we point out that the Navier-Stokes equations with the first-order velocity-slip boundary condition are often misused to predict the apparent gas permeability of the porous media; that is, any nonlinear dependence of the apparent gas permeability with the Knudsen number, predicted from the Navier-Stokes equations, is not reliable. Worse still, for some type of gas-surface interactions, even the ``filtered'' linear dependence of the apparent gas permeability with the Knudsen number is of no practical use since, compared to the numerical solution of the Bhatnagar-Gross-Krook equation, it is only accurate when the ratio between the apparent and intrinsic permeabilities is $\lesssim1.5$.
 
\end{abstract}


\section{Introduction}

Unconventional gas reservoirs have recently received significant attention due to the shale gas revolution in North America~\citep{Wang2014Shale}, but the accurate prediction of unconventional gas production remains a grand research challenge. The permeability of the porous medium is an important parameter in the unconventional gas industry, as it describes how fast the gas can be extracted. The permeability at the representative elementary volume scale can be plugged into macroscopic upscaling equations to predict the gas production~\citep{Javadpour2012,Lunati2014}. Therefore, it is necessary to investigate how the permeability changes with gas pressure, properties of the porous media, and gas-surface interactions.

For laminar flows in highly permeable porous media, Darcy's law states that the flux $\text{q}$ (i.e. discharge per unit area, with units of length per time) is proportional to the pressure gradient $\nabla{p}$:
\begin{equation}\label{Darcy}
\textbf{q}=-\frac{k_\infty}{\mu}\nabla{p},
\end{equation}
where $\mu$ is the shear viscosity of the fluid and $k_\infty$ is the permeability of a porous medium that is independent of the fluid. For this reason, $k_\infty$ is known as the intrinsic permeability.

For gas flows in low permeable porous media, however, the permeability measured from experiments is larger than the intrinsic permeability and increases with the reciprocal mean gas pressure $1/\bar{p}$~\citep{Klinkenberg1941,HeriotWatt2016}. In order to distinguish it from the intrinsic permeability, the term of apparent gas permeability (AGP) $k_a$ has been introduced, which is expressed by~\cite{Klinkenberg1941} as:
\begin{equation}\label{Klinkenberg}
{k_a}={k_\infty}\left(1+\frac{b}{\bar{p}}\right),
\end{equation} 
where $b$ is  the correction factor. 


Darcy's law can be derived from the Navier-Stokes equations (NSEs). The variation of the AGP with respect to the reciprocal mean gas pressure is due to the rarefaction effects, where infrequent collisions between gas molecules not only cause the gas slippage at the solid surface, but also modify the constitutive relation between the stress and strain-rate~\citep{henning}, so that NSEs have to be modified, or completely fail. The extent of rarefaction is characterized by the Knudsen number $Kn$ (i.e. the ratio of the mean free path $\lambda$ of gas molecules to the characteristic flow length $L$):
\begin{equation}\label{meanfreepath}
Kn=\frac{\lambda}{L}, \\~\\ \textrm{and} \\~\\  \lambda=\frac{\mu(T_0)}{\bar{p}}\sqrt{\frac{\pi{RT_0}}{2}},
\end{equation}
where $\mu(T_0)$ is the shear viscosity of the gas at a reference temperature $T_0$, and $R$ is the gas constant. Gas flows can be classified into four regimes\footnote{Note that this partition of flow regime is roughly true for the gas flow between two parallel plates with a distance $L$; for gas flows in porous media, the region of $Kn$ for different flow regimes may change.}: continuum flow $(Kn\lesssim0.001)$ in which NSEs is applicable; slip flow $(0.001<Kn\lesssim0.1)$ where NSEs with appropriate velocity-slip/temperature-jump boundary conditions may be used; transition flow $(0.1<Kn\lesssim10)$ and free-molecular flow $(Kn>10)$, where NSEs break down and gas kinetic equations such as the fundamental Boltzmann equation are used to describe rarefied gas flows~\citep{CE}.

Through systematical experimental measurements, Klinkenberg found that the linear relation~\eqref{Klinkenberg} between the AGP and reciprocal mean gas pressure (or equivalently, the Knudsen number) is only an approximation, as it became ``evident from Tables 5, 6, and 7, the value of constant $b$ increases with increasing pressure'', that is, $b$ decreases when $Kn$ increases. In a latest experiment, a stronger variation of $b$ with the pressure is observed, see Fig. 7 in~\cite{HeriotWatt2016}.

 \begin{figure}
 	\centering
 	{	\includegraphics[scale=0.33,viewport=40 20 790 545,clip=true]{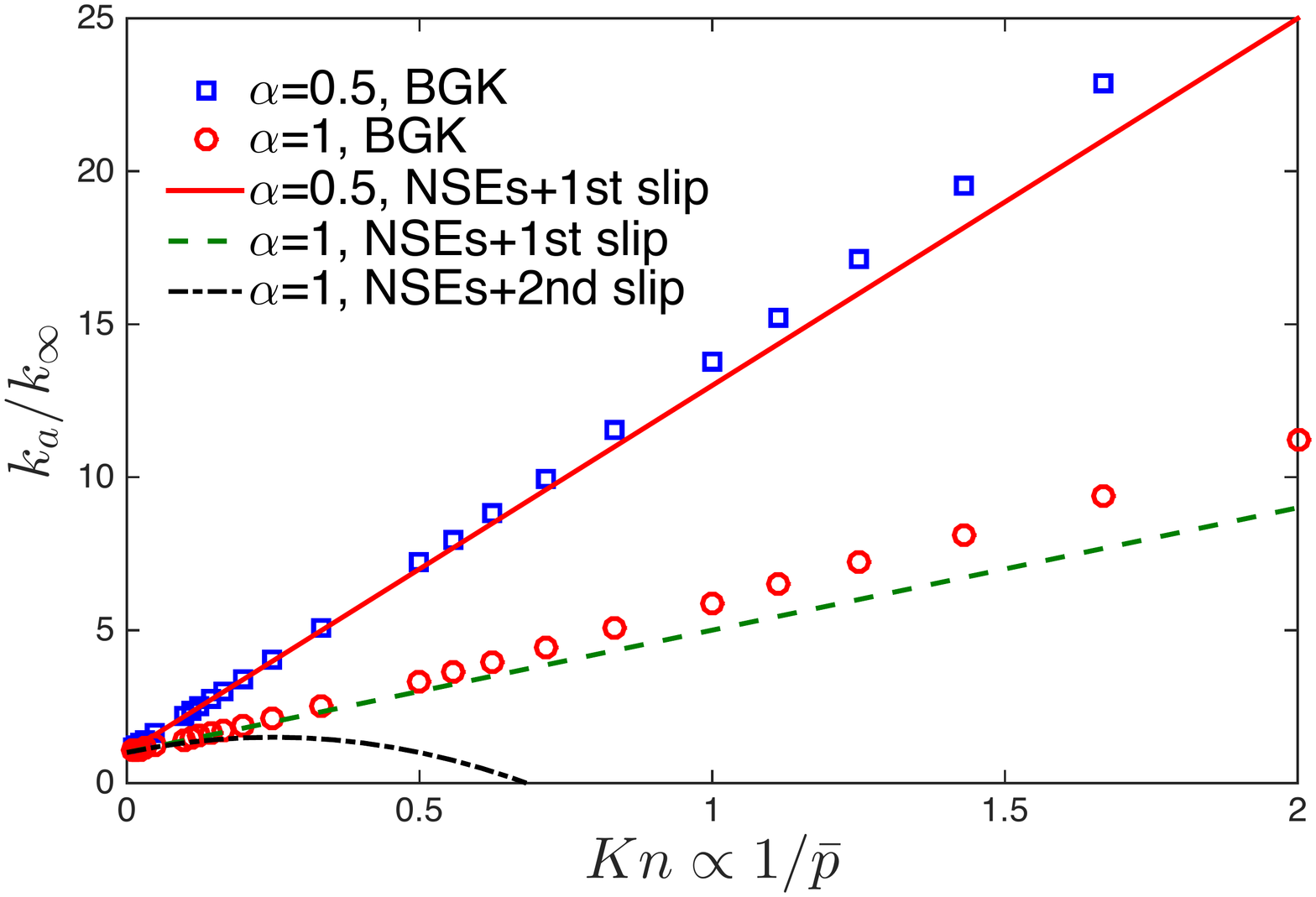}}
 	\caption{The AGP versus the Knudsen number for a gas flow in a straight cylindrical tube, obtained from the numerical simulation of the BGK equation, where $\alpha$ is the tangential momentum accommodation coefficient in the kinetic boundary condition for the gas-surface interaction (see \S~\ref{Num_BGK}). Analytical solutions of the NSEs with first-order and second-order velocity slip boundary conditions are also shown, where the viscous velocity slip coefficients are obtained from~\cite{Loyalka1975} and~\cite{Gibelli_2ndSLIP2011}.}
 	\label{ApparentTube}
 \end{figure}
 
Theoretically, by numerically solving the gas kinetic equation on the exact samples used by Klinkenberg, the nonlinear dependence between the AGP and Knudsen number can be recovered in the whole range of gas pressure. However, due to the complexity of numerical simulations, the porous medium is often simplified by a single straight cylindrical tube (or a bundle of straight tubes with the same radius), as initially suggested by~\cite{Klinkenberg1941}. Then, using the Maxwell's diffuse-specular boundary condition for the gas-surface interaction~\citep{Maxwell1879}, the AGP of a straight tube can be obtained numerically; the numerical solution is often fitted analytically, for example for the diffuse boundary condition,  by the following expression~\citep{Beskok1999,Civan2010}:
\begin{equation}\label{Beskok}
\frac{k_a}{k_\infty}=\left[ 1+\frac{128}{15\pi^2} \tan^{-1}\left(4Kn^{0.4}\right)Kn\right]\left(1+\frac{4Kn}{1+4Kn}\right),
\end{equation}
to predict the unconventional gas production.

So far, although many attempts have been made to predict the AGP, none of them were able to explain the simple experiment by Klinkenberg 76 years ago, in a wide range of gas pressures. Fig.~\ref{ApparentTube} shows the numerical solutions of the Bhatnagar-Gross-Krook (BGK) kinetic equation (see~\S~\ref{Num_BGK}), as well as that of NSEs with velocity-slip boundary conditions. It is clear that both the BGK equation and NSEs with  first-order velocity-slip boundary condition (FVBC) cannot explain  Klinkenberg's experimental results that ``the correction factor decreases when the Knudsen number increases''. In fact, NSEs give a constant correction factor, while the correction factor from the BGK equation increases with the Knudsen number. NSEs with second-order velocity-slip boundary condition~\citep{Beskok_book,HeriotWatt2016} seem to produce the same trend as observed in Klinkenberg's experiments at small $Kn$, but they quickly predict a negative AGP as $Kn$ increases, which is definitely incorrect. In addition, in the experiment the AGP could increase about 30 times, but the use of second-order velocity-slip boundary condition predicts a maximum AGP that is only 1.5 times the intrinsic permeability.

%

%

The accuracy of gas kinetic equations in modeling rarefied gas flows has been tested extensively in various fields from aerodynamics to microfluidics. We believe that one of the reasons for the contradiction between the solution of the BGK equation and the experimental results is the use of oversimplified straight tubes, where the streamlines are straight instead of tortuous in the experiment. The main objective of the present work is to conduct numerical simulations on simplified porous media (which are much simpler than the real samples to make the calculation tractable but more complicated than straight tubes) to pinpoint factors that lead to the same relation (trend) between the correction factor and Knudsen number, as initially observed by~\cite{Klinkenberg1941} and confirmed in many other experiments.

It should be noted that in a recent work, based on the NSEs with FVBC, \cite{Didier2016} found that the AGP of the porous medium is not only a nonlinear function of $Kn$, but also has the similar trend as observed in~\cite{Klinkenberg1941}. This result, however, is highly questionable, because the NSEs were used beyond their validity. Through our theoretical analysis and numerical calculations, we show that the NSEs with FVBC can only predict the AGP of the porous medium to the first-order accuracy of $Kn$. This forms the second objective of this paper, as the NSEs with FVBC are widely misused to predict the AGP of porous media.

\begin{figure}
	\centering
	{	\includegraphics[scale=0.55,viewport=140 270 390 525,clip=true]{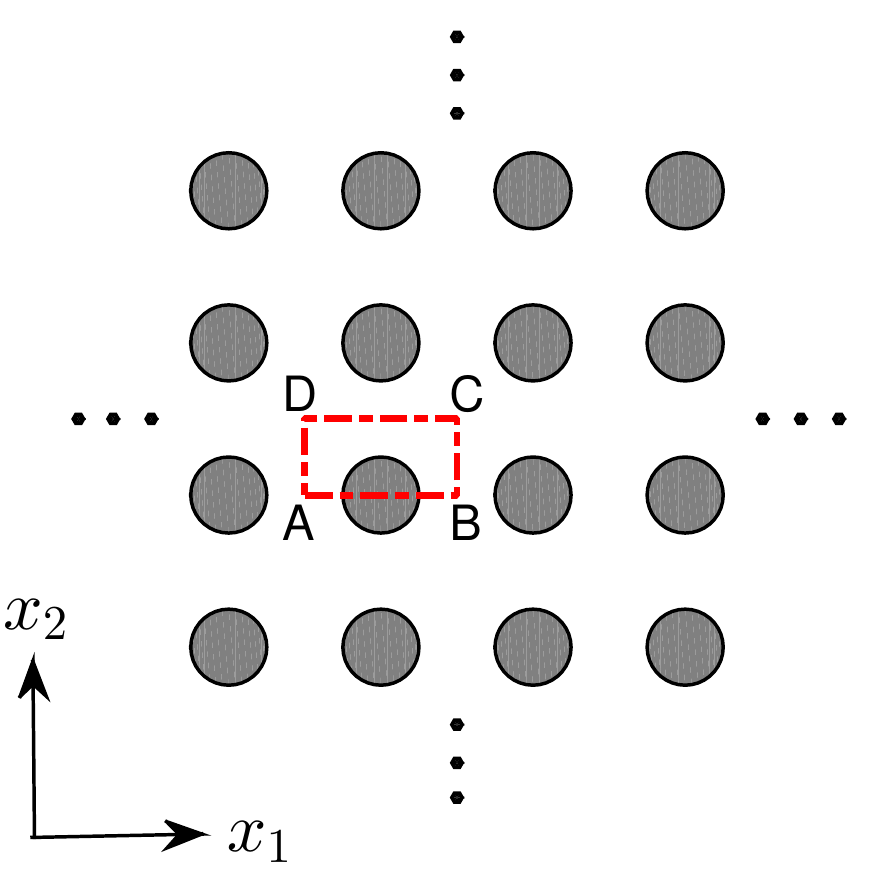}}
	\caption{A two-dimensional porous medium consisting of a periodic array of discs. A, B, C, and D are the four corners of the unit rectangular cell (computational domain used below). The length of the side AB is $L$, while that of AD is $L/2$. Other complex porous media can be generated by adding random solids into the rectangle ABCD, and applying the periodic boundary condition at sides AD and BC, but the symmetrical boundary condition along sides AB and CD, respectively.}
	\label{GEO}
\end{figure}

The remainder of the paper is organized as follows: In \S~\ref{Sec2} we introduce the mesoscopic BGK kinetic model and the gas kinetic boundary condition, as well as the macroscopic regularized 20-moment (R20) equations to describe rarefied gas flows in porous media. We also analyze the limitation of the NSEs and FVBC. In \S~\ref{Num_simulaiton}, numerical simulations are performed to analyze the variation of the AGP with the Knudsen number and to assess the validation of the NSEs with FVBC. The influence of the gas-surface interaction on the AGP is analyzed, and possible factors that lead to the observation of~\cite{Klinkenberg1941} are identified in~\S~\ref{Klin_exp}. The paper closes with some final comments in \S~\ref{Conclusions}.


\section{State of the problem}\label{Sec2}

Consider a gas flowing through a periodic porous medium. Suppose the geometry along the $x_3$-direction is uniform and infinite, the gas flow is effectively two-dimensional and can be studied in a unit rectangular cell ABCD, with appropriate governing equations and boundary conditions. One example of the porous medium consisting of a periodic array of discs is shown in Fig.~\ref{GEO}; other more complex porous media can be generated by adding more solids randomly in the unit rectangular cell ABCD. We are interested in how the AGP varies with the Knudsen number, properties of porous media,  and gas-surface interaction.

\subsection{The mesoscopic description: the gas kinetic theory}\label{Num_BGK}


The Boltzmann equation, which is fundamental in the study of rarefied gas dynamics from the continuum to the free-molecular flow regimes, uses the  distribution function $f(t,\textbf{x},\textbf{v})$ to describe the system state:
\begin{equation}\label{Boltzmann}
\frac{\partial f}{\partial t}+v_1\frac{\partial f}{\partial x_1}+v_2\frac{\partial f}{\partial x_2}+v_3\frac{\partial f}{\partial x_3}=\mathcal{C}(f),
\end{equation}
where $\textbf{v}=(v_1,v_2,v_3)$ is the three-dimensional molecular velocity normalized by the most probable speed $v_m=\sqrt{2RT_0}$, $\textbf{x}=(x_1,x_2,x_3)$ is the spatial coordinate normalized by the length $L$ of the side AB, $t$ is the time normalized by $L/v_m$, $f$ is normalized by $\bar{p}/v_m^3mRT_0$ with $m$ being the molecular mass, while $\mathcal{C}$ is the Boltzmann collision operator. In order to save the computational cost, $\mathcal{C}$ is usually replaced by the relaxation-time approximation~\citep{Bhatnagar1954}, resulting in the BGK equation.

When the porous medium is so long that the pressure gradient is small, namely,  $|Ldp/pdx_1|\ll1$ with $p$ being the local gas pressure, the BGK equation can be linearized. The distribution function is expressed as $f=f_{eq}(1+h)$, where the equilibrium distribution function is defined as
\begin{equation}
f_{eq}=\frac{\exp(-|\textbf{v}|^2)}{\pi^{{3}/{2}}},
\end{equation}  
and the perturbation $h$ is governed by:
\begin{equation}\label{BGK}
\frac{\partial h}{\partial t}+v_1\frac{\partial h}{\partial x_1}+v_2\frac{\partial h}{\partial x_2}=\frac{\sqrt{\pi}}{2Kn}\left[\varrho+2u_1v_1+2u_2v_2+\tau\left(|\textbf{v}|^2-\frac{3}{2}\right)-h\right],
\end{equation}
with macroscopic quantities such as the perturbed number density $\varrho$ of gas molecules, the velocity $u_1$ and $u_2$, and the perturbed temperature $\tau$ being calculated as
\begin{equation}
\begin{split}
\varrho=\int{h}f_{eq}d\textbf{v},\ \ \ \ 
(u_1,u_2)=\int(v_1,v_2){h}f_{eq}d\textbf{v},\ \ \ \ 
\tau=\frac{2}{3}\int |\textbf{v}|^2{h}f_{eq}d\textbf{v}-\varrho.
\end{split}
\end{equation}

The kinetic equation~\eqref{BGK} has to be supplied with the boundary condition. Suppose the pressure gradient is along the $x_1$ direction, on the inlet and outlet of the computational domain ABCD (the coordinates of the four corners A, B, C, and D are $(-0.5,0), (0.5,0), (0.5, 0.5)$, and $(-0.5,0.5)$, respectively), the pressure gradient is applied and  periodic condition for the flow velocity is used~\citep{GraurVacuum2012}:
\begin{equation}
h\left(\mp0.5,x_2,v_1,v_2, v_3\right)=\pm1+h\left(\pm0.5,x_2,v_1,v_2,v_3\right), \text{ when } v_1\gtrless0,
\end{equation} 
at the lines AB and CD, the specular reflection boundary condition is used to account for the spatial symmetry: 
\begin{equation}
\begin{aligned}[b]
h\left(x_1,0,v_1,v_2, v_3\right)=&h\left(x_1,0,v_1,-v_2, v_3\right), \quad \text{when } v_2>0,\\
h\left(x_1,0.5,v_1,v_2, v_3\right)=&h\left(x_1,0.5,v_1,-v_2, v_3\right), \text{when } v_2<0,
\end{aligned}
\end{equation}
while at the solid surface, the diffuse-specular boundary condition is used~\citep{Maxwell1879}:
\begin{equation}\label{diffuse}
h(\textbf{v})=\frac{2\alpha}{\pi} \int_{v_n'<0} |v_n'| h(\textbf{v}')\exp(-|\textbf{v}'|^2) d\textbf{v}'+(1-\alpha)h(\textbf{v}-2\textbf{n}v_n),
\end{equation}
where $\textbf{n}$ is the outer normal vector of the solid surface, and $\alpha$ is the tangential momentum accommodation coefficient (TMAC). This boundary condition assumes that, after collision with the surface, a gas molecule is specularly reflected with probability $1-\alpha$, otherwise it is reflected diffusely (i.e. reflected towards every direction with equal probability, in a Maxwellian velocity distribution). Purely diffuse or specular reflections take place for $\alpha=1$ or $\alpha=0$, respectively. Note that most of the current studies, including the widely used equation~\eqref{Beskok}, use the diffuse boundary condition.


The AGP, which is normalized by $L^2$, is calculated by
\begin{equation}\label{apparentP}
k_a=2\sqrt{\frac{1}{\pi}}KnG_P, 
\end{equation}
where $G_p=2\int_0^{1/2}u_1(x_2)dx_2$ is the dimensionless mass flow rate.

\subsection{The macroscopic description: NSEs and moment equations}\label{macro_equation}

Historically, the state of a gas is first described by macroscopic quantities such as the density $\rho$, velocity $u_i$, and temperature $T$; and its dynamics is described by the Euler equations or NSEs (based on the empirical Newton's law for stress and Fourier's law for heat flux). These equations, however, can be derived rigorously from the Boltzmann equation, at various order of approximations.

By taking the velocity moments of the Boltzmann equation~\eqref{Boltzmann}, the five macroscopic quantities are governed by the following equations: 
\begin{eqnarray}
\frac{\partial \rho}{\partial t}+\frac{\partial \rho {u_i}}{\partial {x_i}} = 0, \label{MASS} \\
\frac{\partial \rho u_i}{\partial t}+\frac{{\partial \rho {u_i}{u_j}}}{{\partial {x_j}}} + \frac{{\partial {\sigma _{ij}}}}{{\partial {x_j}}} =  - \frac{{\partial p}}{{\partial {x_i}}}, \label{MOMENTUM}\\
\frac{{\partial \rho T}}{{\partial t}} + \frac{{\partial \rho {u_i}T}}{{\partial {x_i}}} + \frac{2}{{3R}}\frac{{\partial {q_i}}}{{\partial {x_i}}} =  - \frac{2}{{3R}}\left( {p\frac{{\partial {u_i}}}{{\partial {x_i}}} + {\sigma _{ij}}\frac{{\partial {u_j}}}{{\partial {x_i}}}} \right). \label{ENERGY}
\end{eqnarray}

However, the above equations are not closed, since expressions for the shear stress $\sigma_{ij}$ and heat flux $q_i$ are not known. One way to close~\eqref{MASS}-\eqref{ENERGY} is to use the Chapman-Enskog expansion, where the distribution function is expressed in the power series of $Kn$~\citep{CE}:
\begin{equation}
f=f^{(0)}+Kn f^{(1)}+Kn^2 f^{(2)}+\cdots,
\end{equation}
where $f^{(0)}$ is the equilibrium Maxwellian distribution function. When $f=f^{(0)}$, we have $\sigma_{ij}=0$ and $q_i=0$, and~\eqref{MASS}-\eqref{ENERGY} reduce to the Euler equations. When the distribution function is truncated at the first-order of $Kn$, that is, 
\begin{equation}\label{NS_fist}
f=f^{(0)}+Kn f^{(1)},
\end{equation}
we have 
\begin{equation}\label{GTMNSF}
\sigma_{ij} =-2\mu \frac{\partial u_{<i}}{\partial x_{j>}} \\~\\ \textrm{and} \\~\\  q_i = -\frac{15}{4}R\mu \frac{\partial T}{\partial x_i},
\end{equation}
where the angular brackets are used to denote the traceless part of a symmetric tensor, so that  \eqref{MASS}-\eqref{ENERGY} reduce to NSEs. When $f=f^{(0)}+Kn f^{(1)}+Kn^2 f^{(2)}$, Burnett equations can be derived but are not used nowadays due to their intrinsic instabilities~\citep{Colin2008}.


Alternatively, by combining the moment method  of~\cite{Grad1949} and the Chapman-Enskog expansion, the regularized 13-, 20- and 26-moment equations~\citep{Struchtrup2003,Gu2009} can be derived from the Boltzmann equation to describe rarefied gas flows at different levels of rarefaction~\citep{TorrihonReview2016}. Here R20 equations are used, which, in addition to~\eqref{MASS}-\eqref{ENERGY}, include governing equations for the high-order moments $\sigma_{ij}$, $q_i$, and $m_{ijk}$:
\begin{equation}\label{STRESS}
\frac{\partial \sigma_{ij}}{\partial t}+\frac{{\partial {u_k}{\sigma _{ij}}}}{{\partial {x_k}}} + \frac{{\partial {m_{ijk}}}}{{\partial {x_k}}} = - \frac{p}{\mu }{\sigma _{ij}} - 2p\frac{{\partial {u_{ < i}}}}{{\partial {x_{j > }}}}  - \frac{4}{5}\frac{{\partial {q_{ < i}}}}{{\partial {x_{j > }}}} - 2{\sigma _{k < i}}\frac{{\partial {u_{j > }}}}{{\partial {x_k}}},
\end{equation}
\begin{eqnarray}\label{HFLUX}
\frac{\partial q_i}{\partial t} 
 &+& \frac{{\partial {u_j}{q_i}}}{{\partial {x_j}}} 
 + \frac{1}{2}\frac{{\partial {R_{ij}}}}{{\partial {x_j}}}
 = - \frac{2}{3}\frac{p}{\mu }{q_i}
- \frac{5}{2}p\frac{{\partial RT}}{{\partial {x_i}}} 
+ \frac{{{\sigma _{ij}}}}{\rho }\left( {\frac{{\partial p}}{{\partial {x_j}}} + \frac{{\partial {\sigma _{jk}}}}{{\partial {x_k}}}} \right)  
-  RT\frac{{\partial {\sigma _{ij}}}}{{\partial {x_j}}} \nonumber \\ 
&-& \frac{7}{2}{\sigma _{ij}}\frac{{\partial RT}}{{\partial {x_j}}} 
- \left( {\frac{2}{5}{q_i}\frac{{\partial {u_j}}}{{\partial {x_j}}} + \frac{7}{5}{q_j}\frac{{\partial {u_i}}}{{\partial {x_j}}} + \frac{2}{5}{q_j}\frac{{\partial {u_j}}}{{\partial {x_i}}}} \right) - {m_{ijk}}\frac{{\partial {u_j}}}{{\partial {x_k}}} - \frac{1}{6}\frac{{\partial \Delta }}{{\partial {x_i}}},
\end{eqnarray}
\begin{eqnarray}\label{MIJK}
\frac{\partial m_{ijk}}{\partial t} &+& \frac{{\partial {u_l}{m_{ijk}}}}{{\partial {x_l}}} + \frac{{\partial {\phi _{ijkl}}}}{{\partial {x_l}}} =
 - \frac{3}{2}\frac{p}{\mu }{m_{ijk}} - 3\frac{{\partial RT{\sigma _{ < ij}}}}{{\partial {x_{k > }}}}
 - \frac{{12}}{5}{q_{ < i}}\frac{{\partial {u_j}}}{{\partial {x_{k > }}}} \nonumber \\
 &+& 3\frac{{{\sigma _{ < ij}}}}{\rho }\left( {\frac{{\partial p}}{{\partial {x_{k > }}}} + \frac{{\partial {\sigma _{k > l}}}}{{\partial {x_l}}}} \right).
\end{eqnarray}

The constitutive relationships between the unknown higher-order moments ($R_{ij}$, $\Delta$ and $\phi_{ijkl}$) and lower-order moments were given by Structrup \& Torrilhon (2003) and Gu \& Emerson (2009) to close~\eqref{MASS} to~\eqref{MIJK}. For slow flows in porous media, it is adequate to use the gradient transport terms only~\citep{Taheri2009,GuPRE2010} and they are:
\begin{equation}
{\phi _{ijkl}} =  - \frac{{4\mu }}{{{C_1}\rho }}\frac{{\partial {m_{ < ijk}}}}{{\partial {x_{l > }}}}, \ \ \ \ 
{R_{ij}} =  - \frac{24}{5}\frac{\mu}{p} RT\frac{{\partial {q_{ < i}}}}{{\partial {x_{j > }}}}, \ \ \ \ 
\Delta  =  - 12\frac{\mu }{{p}} RT\frac{{\partial {q_k}}}{{\partial {x_k}}},
\end{equation}
where the collision constant $C_1$ is  2.097 for Maxwell molecules~\citep{Gu2009}.

Macroscopic wall boundary conditions were obtained from the diffuse-specular boundary condition~\citep{Maxwell1879}. In a frame where the coordinates are attached to the wall, with $n_i$ the normal vector of the wall pointing towards the gas and $\tau_i$ the tangential vector of the wall, the velocity-slip parallel to the wall $u_{\tau}$ and temperature-jump conditions are:
\begin{eqnarray}
{u_\tau } =  - \frac{{2 - \alpha }}{\alpha }\sqrt {\frac{{\pi RT}}{2}} \frac{{{\sigma _{n\tau }}}}{{{p_\alpha }}} - \frac{{5{m_{nn\tau }} + 2{q_\tau }}}{{10{p_\alpha }}}, \label{ho_velocity}\\
RT - R{T_w} =  - \frac{{2 - \alpha }}{\alpha }\sqrt {\frac{{\pi RT}}{2}} \frac{{{q_n}}}{{2{p_\alpha }}}  - \frac{{RT{\sigma _{nn}}}}{{4{p_\alpha }}} + \frac{{u_\tau ^2}}{4} - \frac{{75{R_{nn}} + 28\Delta }}{{840{p_\alpha }}} + \frac{{{\phi _{nnnn}}}}{{24{p_\alpha }}},
\end{eqnarray}
where ${p_\alpha } = p + \sigma _{nn}/{2} - (30R_{nn} + 7\Delta)/ 840RT - \phi _{nnnn}/24RT$. The rest of wall boundary conditions for higher-order moments are listed in Appendix~\ref{Wall_hob}. Note that the velocity-slip boundary condition~\eqref{ho_velocity} is also of higher-order due to the appearance of the higher-order moment $m_{nn\tau}$.

Following the above introduction, it is clear that the NSEs with FVBC are only accurate to the first-order of $Kn$; therefore, any AGP predicted from NSEs and showing the nonlinear dependence with $Kn$ is highly questionable. The R20 equations are accurate to the third-order of $Kn$~\citep{henning}, which should give the some AGP of the porous medium as NSEs when $Kn\rightarrow0$, and be more accurate than NSEs as $Kn$ increases. Numerical simulations are also performed in the following section to demonstrate this.

\section{Numerical methods and results}\label{Num_simulaiton}

In this section, we first present analytical/numerical solutions of the NSEs with FVBC, the linearized BGK equation, the R20 equation, as well as the direct simulation Monte Carlo (DSMC) method~\citep{Borner2017}, for rarefied gas flows through porous media, to show the accuracy of the NSEs with FVBC. Then we give a possible explanation to understand Klinkenberg's experimental results.

\subsection{The accuracy of the NSEs and FVBC}\label{accuracy}

We first investigate the rarefied gas flow through a periodic array of discs with diameter $D$, as shown in Fig.~\ref{GEO}. Using the NSEs and FVBC, when the porosity $\epsilon=1-\pi{D}^2/4L^2$ is large, the AGP can be obtained analytically~\citep{Chai2011}:
\begin{equation}\label{per_large_por}
k_a=\frac{1}{8\pi\left(1+2\xi{Kn}\sqrt{\frac{\pi}{\phi}}\right)}\bigg[-\ln\phi-\frac{3}{2}+2\phi-\frac{\phi^2}{2}
+2\xi{Kn}\sqrt{\frac{\pi}{\phi}}\left(-\ln\phi-\frac{1}{2}+\frac{\phi^2}{2}\right)\bigg],
\end{equation}
where $\phi=1-\epsilon$ is the solid fraction and  $\xi$ is related to the viscous velocity slip coefficient. When the Maxwell's diffuse boundary condition~\eqref{diffuse} is used~\citep{Hadjiconstantinou2003slip}, we have $\xi=1.016\sqrt{{4}/{\pi}}=1.15$. The viscous velocity slip coefficient for other values of TMAC can be found in~\cite{Loyalka1975} and~\cite{Gibelli_2ndSLIP2011}. The intrinsic permeability $k_\infty$ is obtained when $Kn=0$.

\begin{figure}
	\centering
	\subfloat[]{\includegraphics[scale=0.6,viewport=30 10 625 285,clip=true]{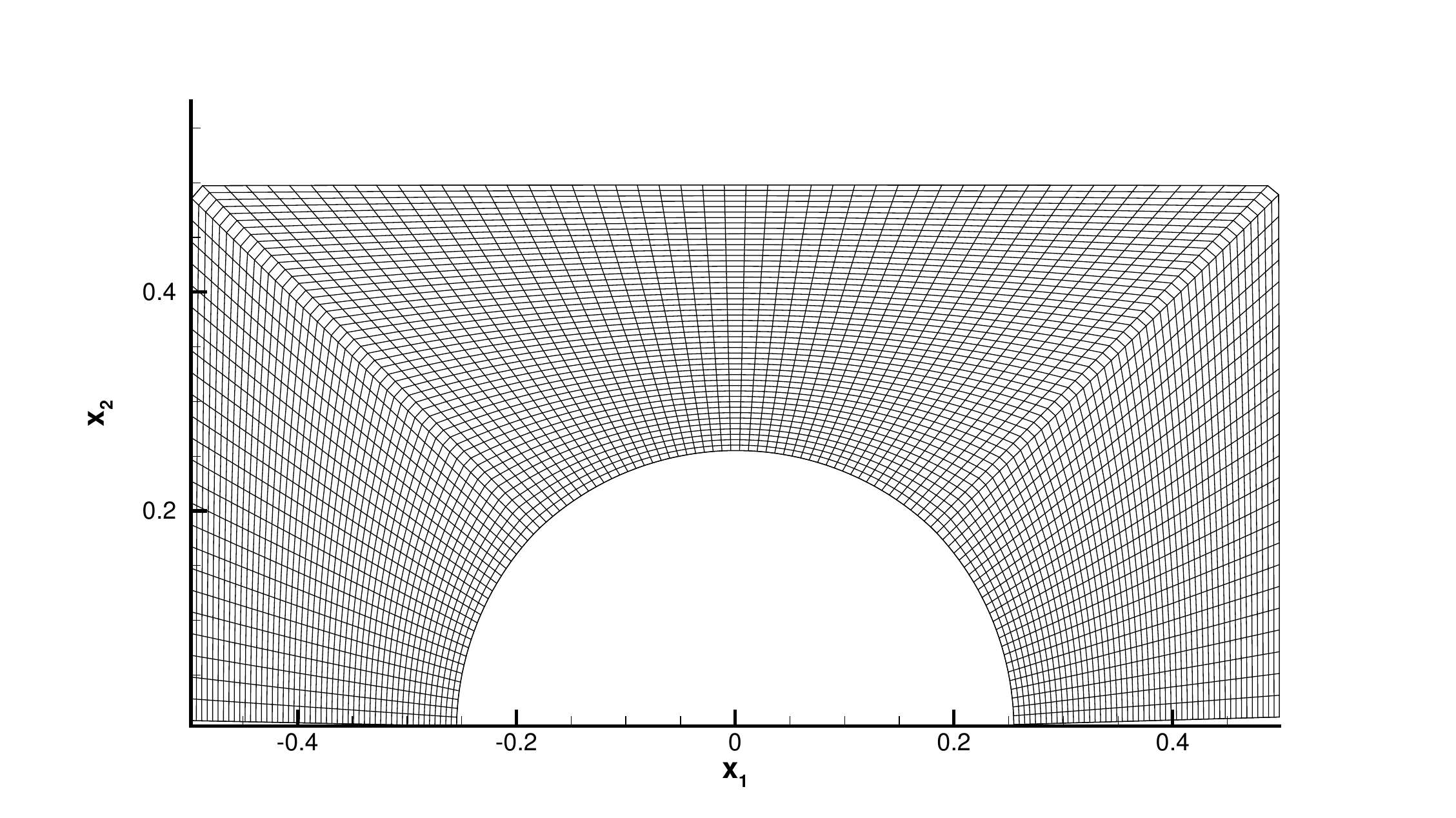}}\\	
	\subfloat[]	{\includegraphics[scale=0.4,viewport=30 30 770 535,clip=true]{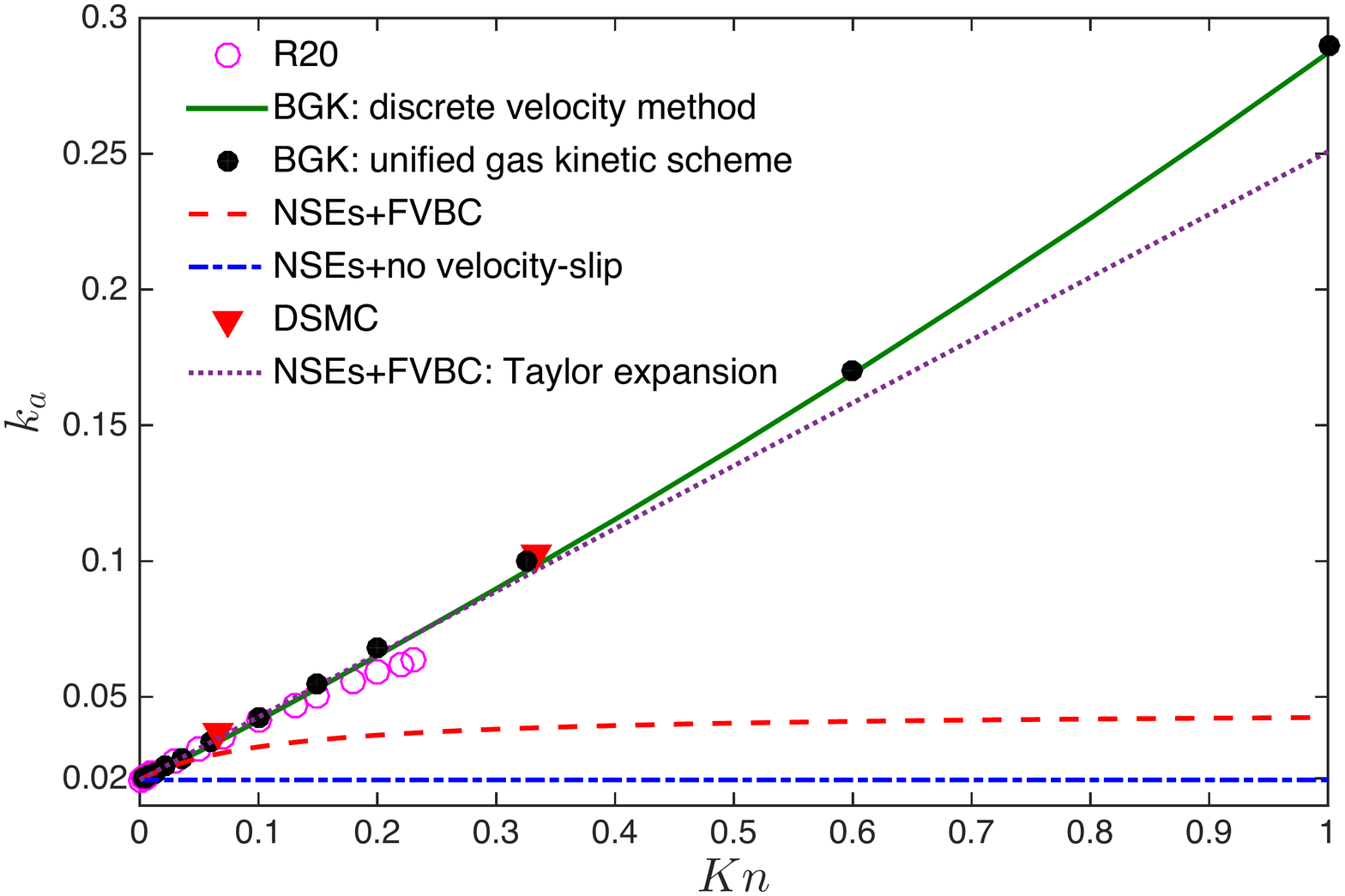}}	
	\caption{(a) The body-fitted mesh used in the unified gas kinetic scheme, when the porosity of the porous medium in Fig.~\ref{GEO} is $\epsilon=0.8$. For clarity, only $100\times50$ cells are shown. (b) The AGP as a function of the Knudsen number, when the diffuse boundary condition is used, i.e. $\alpha=1$ in~\eqref{diffuse}. The solid line and dots are numerical results of the linearized BGK solved by the discrete velocity method and the unified gas kinetic scheme, respectively. The dash-dotted and dashed lines are analytical solutions of NSEs~\eqref{per_large_por}, with the no-slip and first-order velocity-slip boundary conditions, respectively, while the dotted line is the slip-corrected permeability obtained by expanding the analytical solution~\eqref{per_large_por} to the first-order of $Kn$, see~\eqref{per_large_por_taylor}. The DSMC results are obtained from the recent simulation by~\cite{Borner2017}. }\label{Apparent_Permeability}
\end{figure}

For the linearized BGK equation~\eqref{BGK}, two reduced distribution functions were introduced to cast the three-dimensional molecular velocity space into a two-dimensional one, and the obtained two equations are solved numerically by the discrete velocity method~\citep{GraurVacuum2012} and the unified gas kinetic scheme~\citep{Huang2013}. In the unified gas kinetic scheme, a body-fitted structured curvilinear mesh is used, with 150 lines along the radial direction and 300 lines along the circumferential direction, see Fig.~\ref{Apparent_Permeability}(a). In the discrete velocity method, a Cartesian grid with $801\times401$ equally-spaced points is used and the solid surface is approximated by the ``stair-case" mesh. In solving the R20 equations, a body-fitted mesh with $201\times101$ cells is used, and the detailed numerical method is given by~\cite{Gu2009}. The molecular velocity space in the BGK equation is also discretized: $v_1$ and $v_2$ are approximated by the $8\times8$ Gauss-Hermite quadrature when $Kn$ is small ($Kn<0.01$ in this case), and the Newton-Cotes quadrature with $22\times22$ non-uniform discrete velocity points when $Kn$ is large~\citep{lei_Jfm}.

The results for a periodic porous medium with $\epsilon=0.8$ are shown in Fig.~\ref{Apparent_Permeability}. It is seen that both the discrete velocity method on the ``stair-case'' geometry and the unified gas kinetic scheme on the curvilinear mesh have the same results. Also, although we use the linearized BGK model, our numerical results are quite close to those from the DSMC simulation~\citep{Borner2017}. Note that in the DSMC simulation, the Knudsen number and AGP are normalized by the radius and radius square of the disc, respectively, so the data should be re-normalized. Also, the mean free path defined in~\eqref{meanfreepath} is $15\pi/2(7-2\omega)(5-2\omega)$ times larger than that used in DSMC based on the variable hard sphere model, where $\omega=0.81$ for argon~\citep{lei}; so the Knudsen number should be re-scaled.

The accuracy of the  permeability~\eqref{per_large_por} is assessed by comparing to numerical solutions of the BGK and R20 equations. When $Kn\lesssim0.15$, our numerical simulations based on the linearized BGK equation and R20 equations agree with each other, and the AGP is a linear function of $Kn$. When $Kn\gtrsim0.2$, the R20 equations, although being accurate to the third-order of $Kn$, predict lower AGP than that of the BGK equation. For larger Kn beyond the validity of the R20 equations and their boundary conditions, no converged numerical solutions can be obtained. The analytical permeability~\eqref{per_large_por} increases linearly with $Kn$ only when $Kn\lesssim0.02$, and then quickly reaches to a maximum value when $Kn\gtrsim0.2$. This comparison clearly demonstrates that, the NSEs with FVBC are only accurate to the first-order of $Kn$. This result is in accordance with the approximation~\eqref{NS_fist} adopted in the derivation of NSEs from the Boltzmann equation. Although the ``curvature of the solid-gas interface'', claimed by~\cite{Didier2016}, makes the AGP a concave function of $Kn$ in the framework of NSEs with FVBC, higher-order moments in~\eqref{STRESS}-\eqref{MIJK} and the higher-order velocity slip in~\eqref{ho_velocity}, which are derived from the Boltzmann equation and the gas kinetic boundary condition~\eqref{diffuse} to the third-order accuracy of $Kn$, restore linear dependence of the AGP on the Knudsen number when $Kn\lesssim0.15$.

Since theoretically the NSEs with FVBC are only accurate up to the first-order of $Kn$, we expand~\eqref{per_large_por}, a nonlinear function of $Kn$, into the Taylor's series of $Kn$ around $Kn=0$, and retain only the zeroth- and first-order terms; the obtained permeability is called the slip-corrected AGP, and it reads
\begin{equation}\label{per_large_por_taylor}
k_{a,\text{slip}}=\frac{1}{8\pi}\left[-\ln\phi-\frac{3}{2}+2\phi-\frac{\phi^2}{2}+2\xi{Kn}\sqrt{\frac{\pi}{\phi}}\left(1-2\phi+{\phi^2}\right)\right].
\end{equation}
The slip-corrected AGP is plotted as the dotted line in Fig.~\ref{Apparent_Permeability}(b). Surprisingly, this Taylor's expansion, which is valid at small $Kn$, gives a  good agreement with the numerical solution of the linearized BGK equation when $Kn\lesssim0.35$; even in the transition flow regime where the Knudsen number is as high as one, this slip-corrected expression predicts an AGP only about 15\% smaller than the numerical result of the linearized BGK equation. We emphasis here that, however, this is {only a coincidence}; and we will explain this point in \S~\ref{Klin_exp} below.

\begin{figure}
	\centering
	\subfloat[]{\includegraphics[scale=0.6,viewport=30 85 555 375,clip=true]{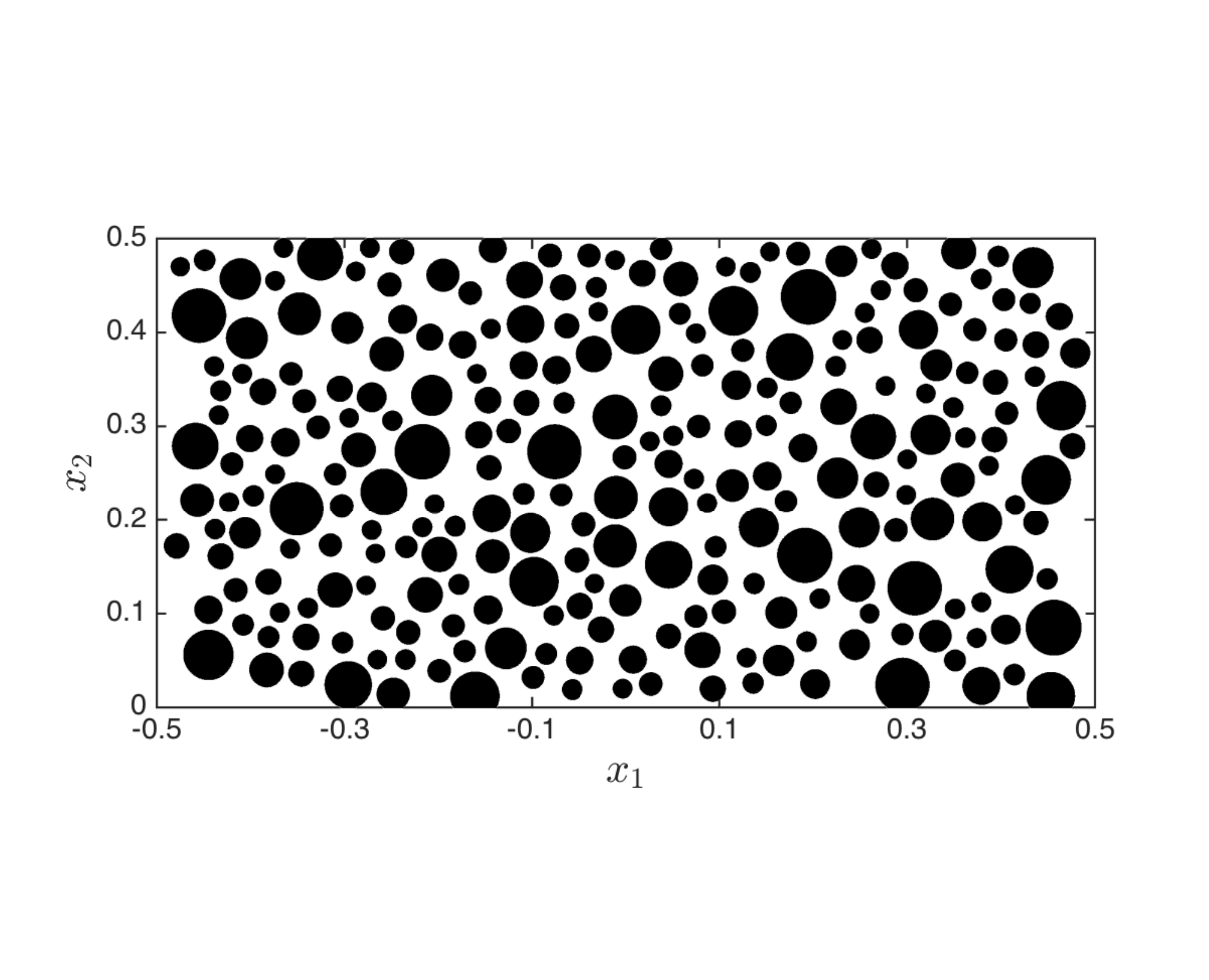}}\\
	\subfloat[]	{\includegraphics[scale=0.35,viewport=40 40 810 510,clip=true]{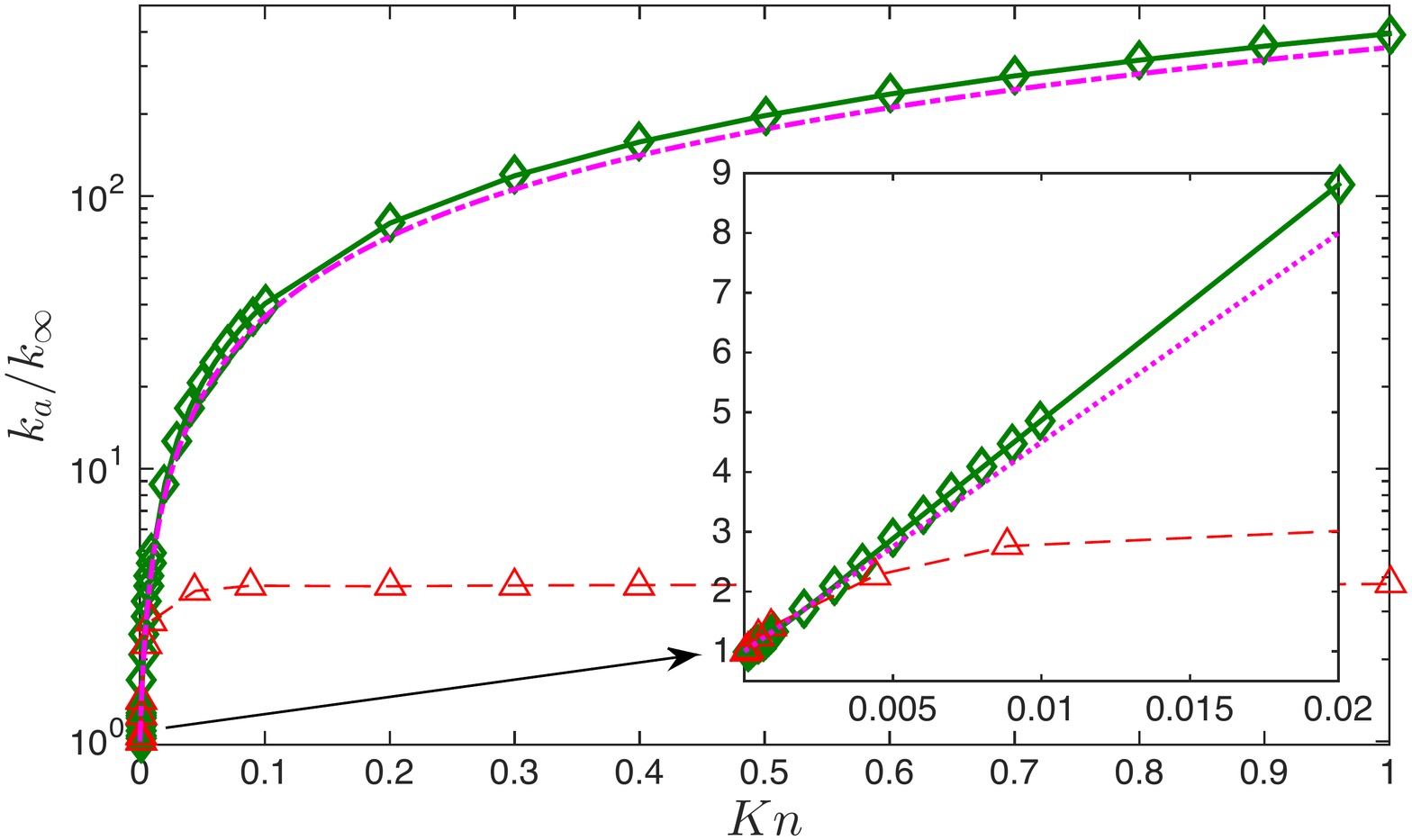}}	
	\caption{(a) The geometry in a unit computational cell, when the porosity is $\epsilon=0.6$. Solids of random size and position are shown in black. The periodic porous medium is generated by placing the whole computational domain inside the unit rectangular cell ABCD in Fig.~\ref{GEO}. $3000\times1500$ cells are used to discretize the spatial domain. (b) The ratio of the AGP to the intrinsic permeability $k_\infty=9.37\times10^{-6}$ as a function of the Knudsen number. The solid and dashed lines are numerical results of the linearized BGK equation and NSEs with FVBC, respectively. The dotted lines are the slip-corrected permeability, obtained by fitting the numerical solution of the NSEs with FVBC at small Knudsen numbers to the first-order of $Kn$. Note that here the effective Knudsen number $Kn^\ast$ defined in~\eqref{effectiveKn} is 73 times of $Kn$.
	}	\label{RandomDisc}
\end{figure}


The conclusion that the NSEs with FVBC is accurate only to the first-order of $Kn$ not only holds for the simple porous medium in Fig.~\ref{GEO}, but also applies to more complex porous media, for example, see the unit cell in Fig.~\ref{RandomDisc}(a) where the porosity is 0.6. In this case, the linearized BGK equation is solved by the discrete velocity method, with a Cartesian mesh of $3000\times1500$ cells; {the grid convergence is verified, as using $6000\times3000$ cells results in only 0.6\% increase of the AGP when $Kn=0.0001$.} The NSEs with FVBC are solved in OpenFOAM using the SIMPLE algorithm and a cell-centered finite-volume discretization scheme, on unstructured grids. A body-fitted computational grid is generated using the OpenFOAM meshing tool, resulting in a mesh of  about 600,000 cells of which the majority are hexahedra and the rest few close to the walls are prisms. Note that the results of R20 equation and the unified gas kinetic scheme are not available because currently the corresponding codes are not yet developed to deal with such a complex geometry.

Due to the small clearance between discs, the characteristic flow length is much smaller than the size of the computational domain $L$. Thus, the effective characteristic flow length $L^\ast$ is often used instead, which is defined as~\citep{Didier2016}:
\begin{equation}
L^\ast=L\sqrt{\frac{12k_\infty}{\epsilon}}, 
\end{equation}
and the effective Knudsen number is
\begin{equation}\label{effectiveKn}
Kn^\ast=\frac{\lambda}{L^\ast}=Kn\sqrt{\frac{\epsilon}{12k_\infty}}.
\end{equation}

From Fig.~\ref{RandomDisc}(b) we see that the AGP obtained from NSEs with FVBC increases linearly when $Kn\lesssim0.002$ (or $Kn^\ast\lesssim0.146$) and then quickly reaches to a constant. Again, the comparison with solution of the linearized BGK equation shows that the NSEs with FVBC are roughly accurate when the AGP is a linear function of $Kn$; in this region of $Kn$, the maximum AGP is only about 1.5 times larger than the intrinsic permeability $k_\infty$. Interestingly, like~\eqref{per_large_por_taylor}, when the numerical solution of the NSEs with FVBC is ``filtered'' to keep only the zeroth- and first-order terms of $Kn$, the resulting solution is only smaller than the numerical solution of the linearized BGK equation by about 15\% in a large range of the gas rarefaction, i.e. $Kn^\ast\lesssim73$, where the AGP is larger than the intrinsic permeability by hundreds of times.

\subsection{The computational time  }


It will be helpful to mention the computational time for the two examples in~\S~\ref{accuracy}. For the discrete velocity method, numerical simulations are performed on Dell workstation (Precision Tower 7910) with Dual processors Intel Xeon CPU E5-2630 v3 2.40GHz. There are 32 cores in total but only 8 cores are used. For simplicity, the $8\times8$ Gauss-Hermite quadrature is used here for all the Knudsen numbers shown in Table~\ref{comput_time}. For the non-uniform velocity grids, the iteration numbers are roughly the same as the Gauss-Hermite quadrature, but the computational time is about 10 times higher due to the large number of discrete velocity points. The AGP is calculated every 1000 iteration steps, and our Fortran programme is terminated when the following convergence criterion
\begin{equation}
\frac{k_a(i)-k_a(i-1000)}{k_a(i)}<10^{-10},
\end{equation}
is reached,  
where $i$ is the iteration step.

\begin{table}
	\centering
	\begin{tabular}{cccrrcrrccccc}
	 \multicolumn{2}{c}{Geometry} & \multicolumn{2}{c}{circular cylinder in Fig.~\ref{Apparent_Permeability}} & & \multicolumn{2}{c}{complex geometry in Fig.~\ref{RandomDisc}}
		\\ 
		\multicolumn{2}{c}{Spatial grid size} & \multicolumn{2}{c}{$801\times401$} & & \multicolumn{2}{c}{$3000\times1500$}
		 \\ 
 $Kn$ & $Kn^\ast$ & iteration steps & time ($s$)  & $Kn^\ast$   & iterations & time ($s$) \\
	$10^{0}$ & $1.85\times10^{0}$ & $2\times10^{3}$ & $68$ & $7.3\times10^{1}$ &$4\times10^{3}$ & $2076$\\
	$10^{-1}$ &$1.85\times10^{-1}$ & $2\times10^{3}$ & $68$ & $7.3\times10^{0}$ &$3\times10^{3}$ & $1558$\\
	$10^{-2}$ &$1.85\times10^{-2}$ & $7\times10^{3}$ & $238$ & $7.3\times10^{-1}$ &$5\times10^{3}$ & $2584$\\
	$10^{-3}$ &$1.85\times10^{-3}$ & $366\times10^{3}$ & $12381$ & $7.3\times10^{-2}$ &$9\times10^{3}$ & $4943$
	\end{tabular}\par  
	\caption{Iteration steps and elapsed time when the linearized BGK equation is solved by the implicit discrete velocity method. The time is measured by wall-clock time.
		 }\label{comput_time}
\end{table}

Generally speaking, when the implicit scheme is used to solve the gas kinetic equations, iteration steps increase when the Knudsen number decreases. From Table~\ref{comput_time} we find that this is true except for the complex geometry where the iteration steps at $Kn=0.1$ are less than that of $Kn=1$. This is because we start the numerical simulation from $Kn=1$; and when the solution is converged, it is used as the initial guess in the computation of $Kn=0.1$, and so on. Thus, it is possible that the solution for smaller $Kn$ requires fewer iteration steps, especially in complicated porous media. Note that for the circular cylinder case we need many iteration steps at $Kn=0.001$; this is because the flow is in the continuum regime where the collision dominates so that the exchange of information due to streaming is extremely slow, and consequently, the convergence is slow. In practical calculations, however, we do not have to calculate the AGP at such a small $Kn$, because it is only about 2\% larger than the intrinsic permeability; we only have to calculate the intrinsic permeability by the Navier-Stokes solvers such as the OpenFOAM and the multiple relaxation time lattice Boltzmann method for simulating flows in complex porous media~\citep{Pan2006}.

The unified gas kinetic scheme solves the linearized BGK equation explicitly, and hence the computational time is high. For example, the iteration steps are 350000 when $Kn=1$. The implicit unified gas kinetic scheme can speed up the convergence by hundreds of times~\citep{zhuyajun2016}, but the code for porous media flow simulations is still under development. This scheme has the unique advantage that the spatial resolution can be coarser than that of the discrete velocity method to reach the same order of accuracy, so that the computational memory usage can be significantly reduced.

The OpenFOAM solver running on the same Dell workstation (using one core) takes about an hour to obtain the converged solution for the complex porous medium case in Fig.~\ref{RandomDisc}, at each Knudsen number.

The computational efforts to solve the moment equations are opposite to that for the BGK kinetic equation. When $Kn$ is small, a solution can be obtained quickly with large iteration relaxation factors~\citep{Tang2013CCP}. As the Knudsen number increases, the values of the relaxation factors have to be reduced and more iterations are required to reach to a converged solution. In the present study in Fig.~\ref{Apparent_Permeability}, single processor of Intel i7-4770 CPU at 3.4GHz is used and the computing times range from minutes to hours depending on the Knudsen number. Computer codes dealing with the complex geometry are still under development. We intend to couple deterministic solvers for moment equations and the linearized BGK equation, so that higher numerical accuracy and efficiency can be achieved simultaneously.

\subsection{Possible explanation of the Klinkenberg's experiment}\label{Klin_exp}

From the numerical simulations in~\S~\ref{accuracy}, the Klinkenberg's finding that the correction factor decreases when $Kn$ increases is not observed, even when the porous media are more complicated than the straight cylindrical tube. Instead, we find that the correction factor increases with $Kn$, which is exactly the same as the rarefied gas flowing through a straight cylindrical tube, see Fig.~\ref{ApparentTube}.

From the analytical solution~\eqref{per_large_por_taylor}, we find that, if we use a non-unitary TMAC (say, $\alpha=0.5$), then $\xi\approx3.23$~\citep{Loyalka1975}, and the correction factor is about three times larger than  that of $\alpha=1$ when $Kn$ is small. If at large Knudsen numbers the AGP of $\alpha=0.5$ is only slightly larger than that of $\alpha=1$ (unfortunately this is not the case for rarefied gas flows between two parallel plates or through a straight cylindrical tube), a decrease of the correction factor when $Kn$ increases may be observed in the numerical simulation of the linearized BGK equation in porous media.

To study the influence of the non-unitary TMAC on the AGP, we investigate the rarefied gas passing through an infinite square array of square cylinders, by replacing the discs in Fig.~\ref{GEO} with squares. The only reason to do this is that the specular reflection in the boundary condition~\eqref{diffuse} can be accurately implemented.

Since we use non-dimensional variables, we rewrite the Klinkenberg's famous equation~\eqref{Klinkenberg} in the following form
\begin{equation}\label{Klinkenberg2}
\frac{k_a}{k_\infty}=1+\frac{b}{\bar{p}}\equiv1+{b'}Kn^\ast,
\end{equation}
and investigate the variation of the ``equivalent correction factor'' $b'$ with respect to the Knudsen number and TMAC. Note that $b'$ is proportional to the correction factor $b$ and the proportionality constant is independent of the Knudsen number (reciprocal mean gas pressure) and TMAC.

\begin{figure}
	\centering
	\subfloat[]{\includegraphics[scale=0.38,viewport=30 285 515 545,clip=true]{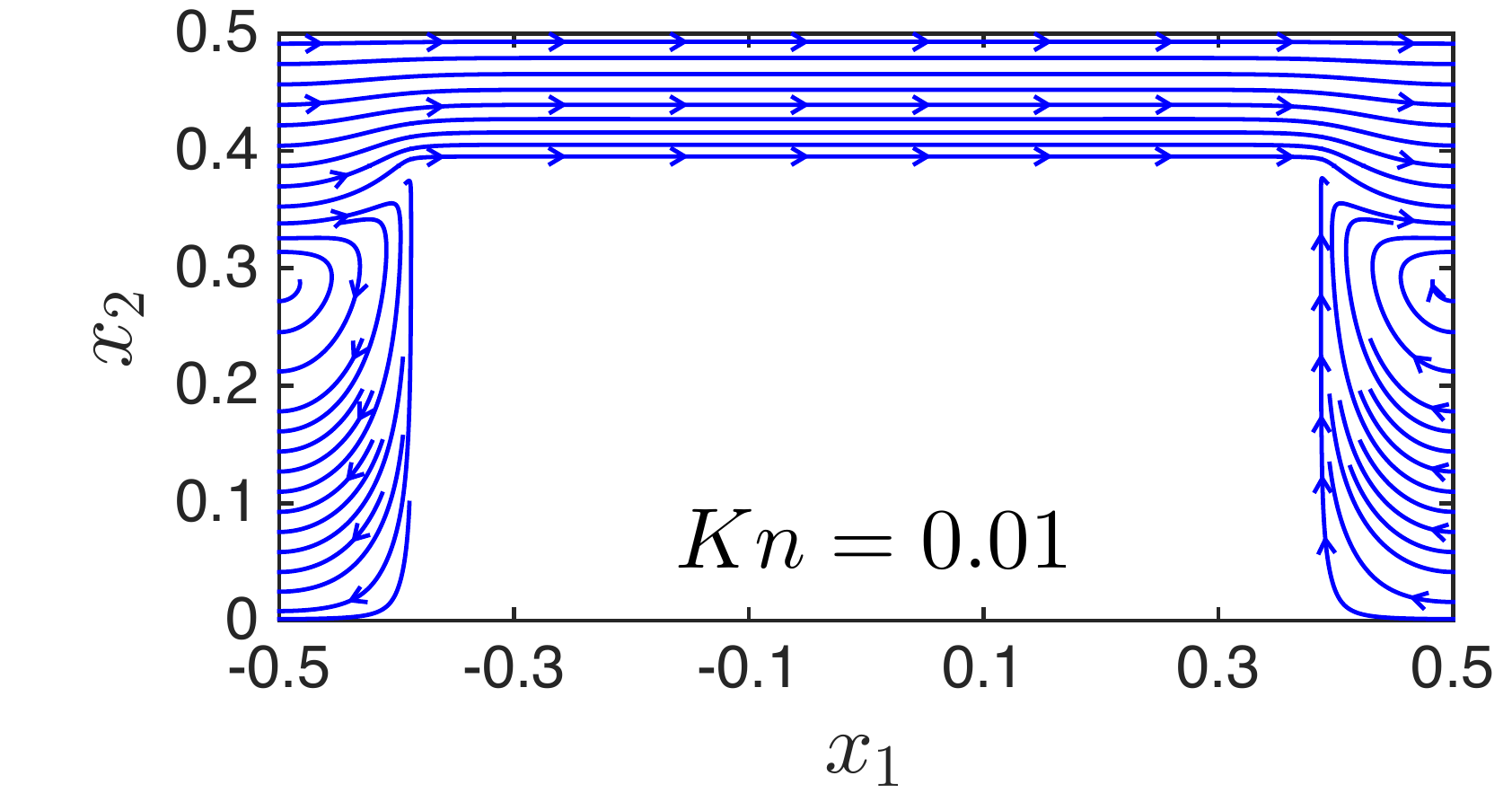}}
	\hskip 0.25cm
	\subfloat[]{\includegraphics[scale=0.38,viewport=30 285 515 545,clip=true]{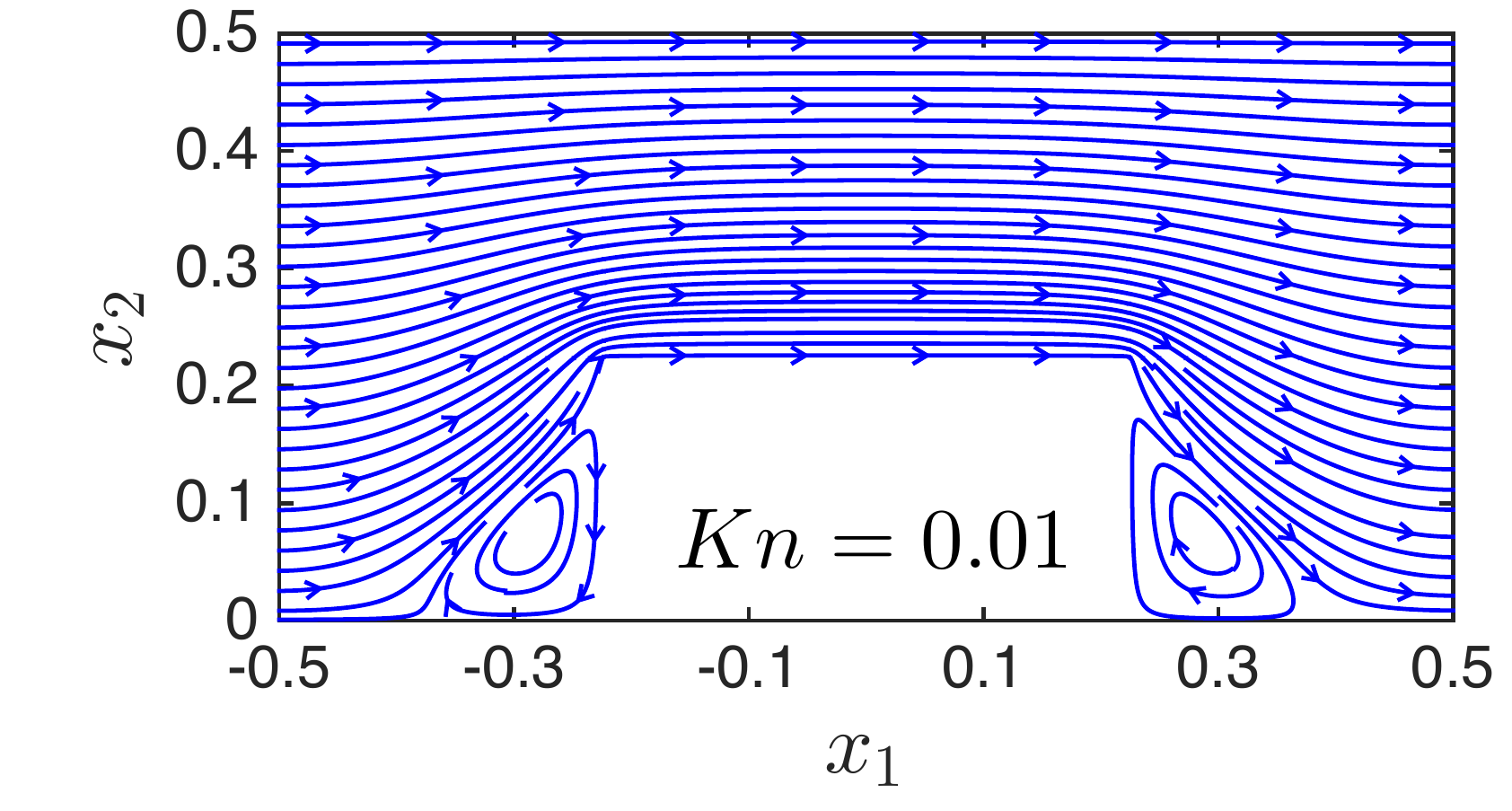}}\\	
	\subfloat[]	{\includegraphics[scale=0.33,viewport=0 200 550 630,clip=true]{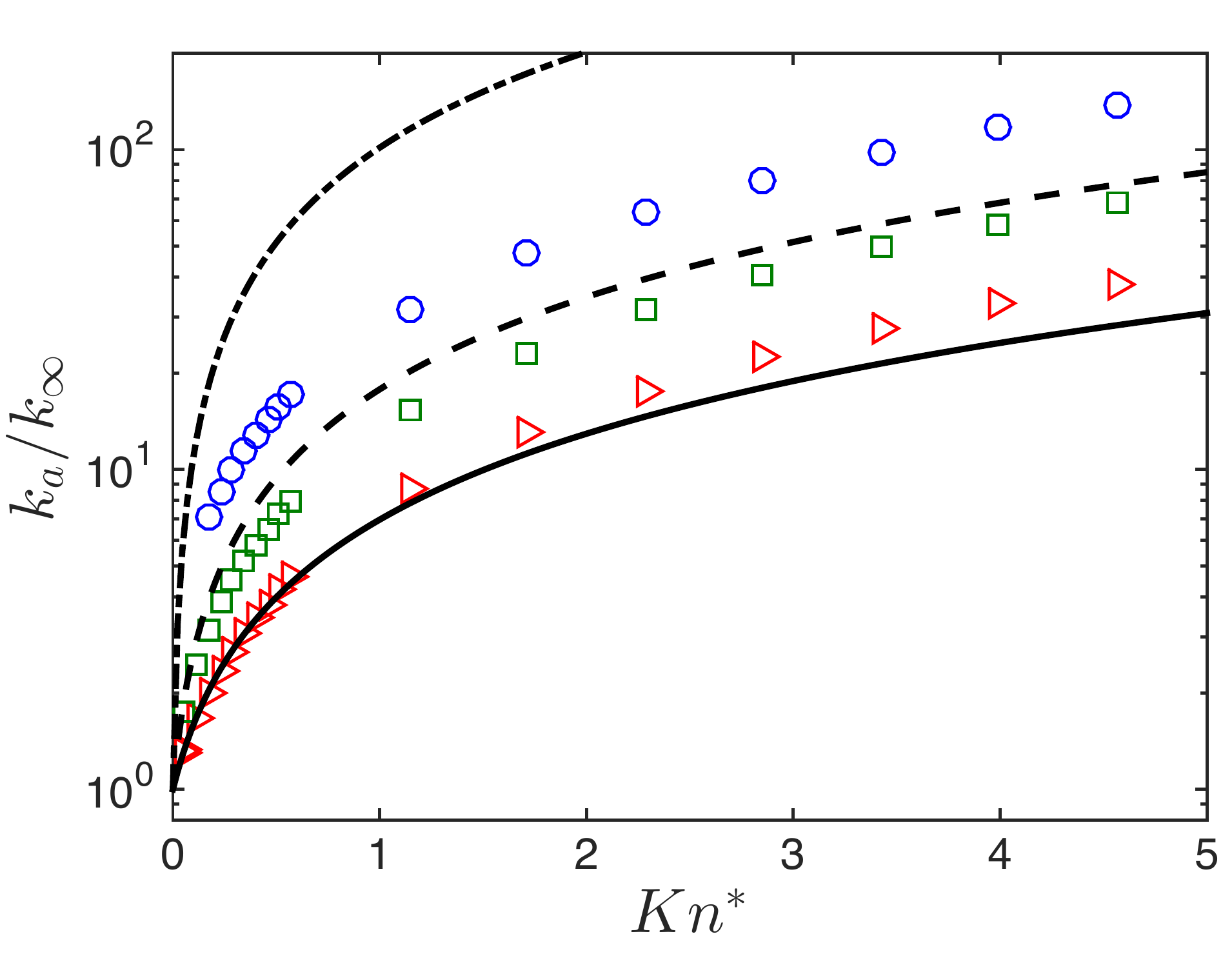}}
	\hskip 0.5cm
	\subfloat[]	{\includegraphics[scale=0.33,viewport=0 200 550 630,clip=true]{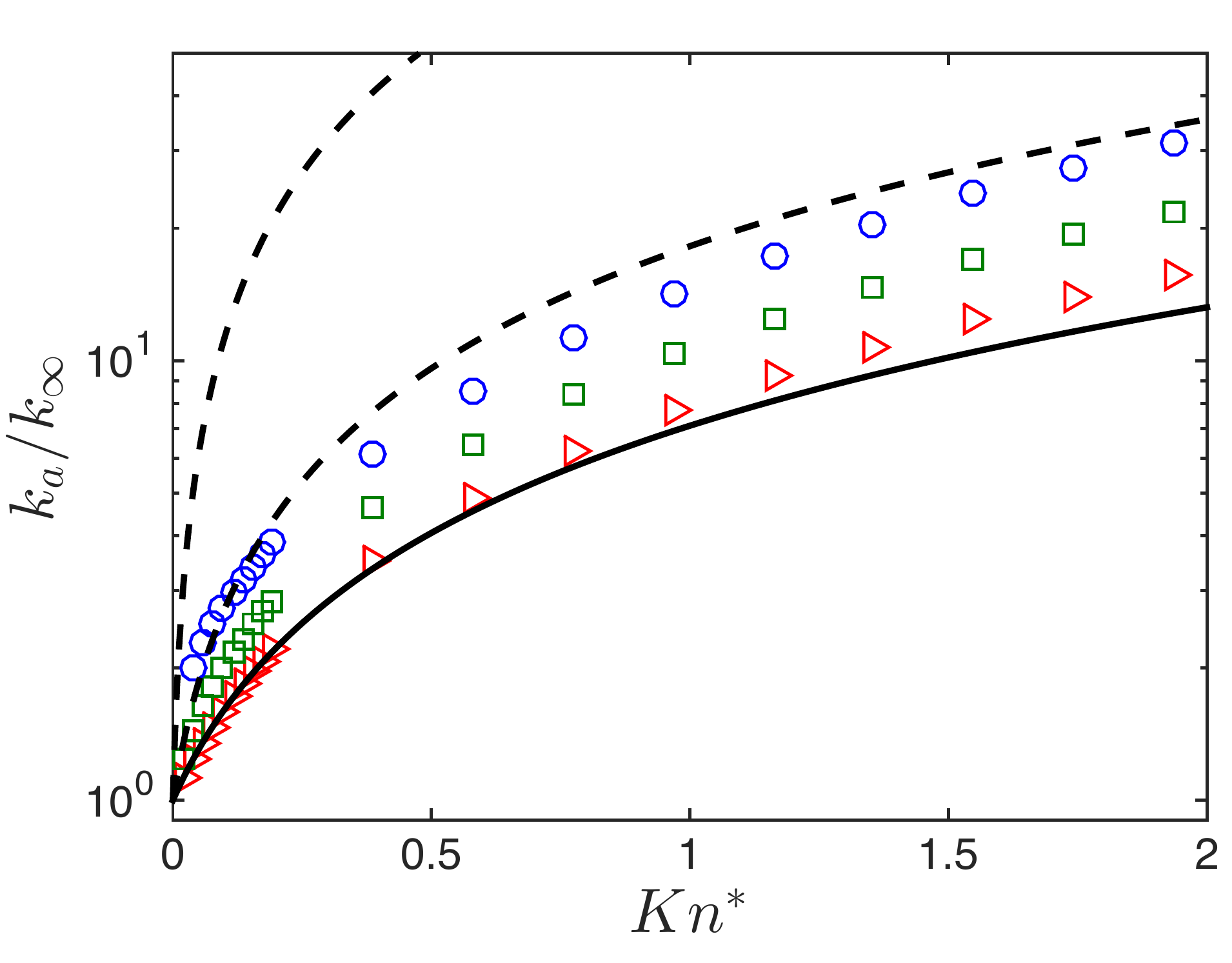}}\\	
	\subfloat[]	{\includegraphics[scale=0.33,viewport=0 200 550 620,clip=true]{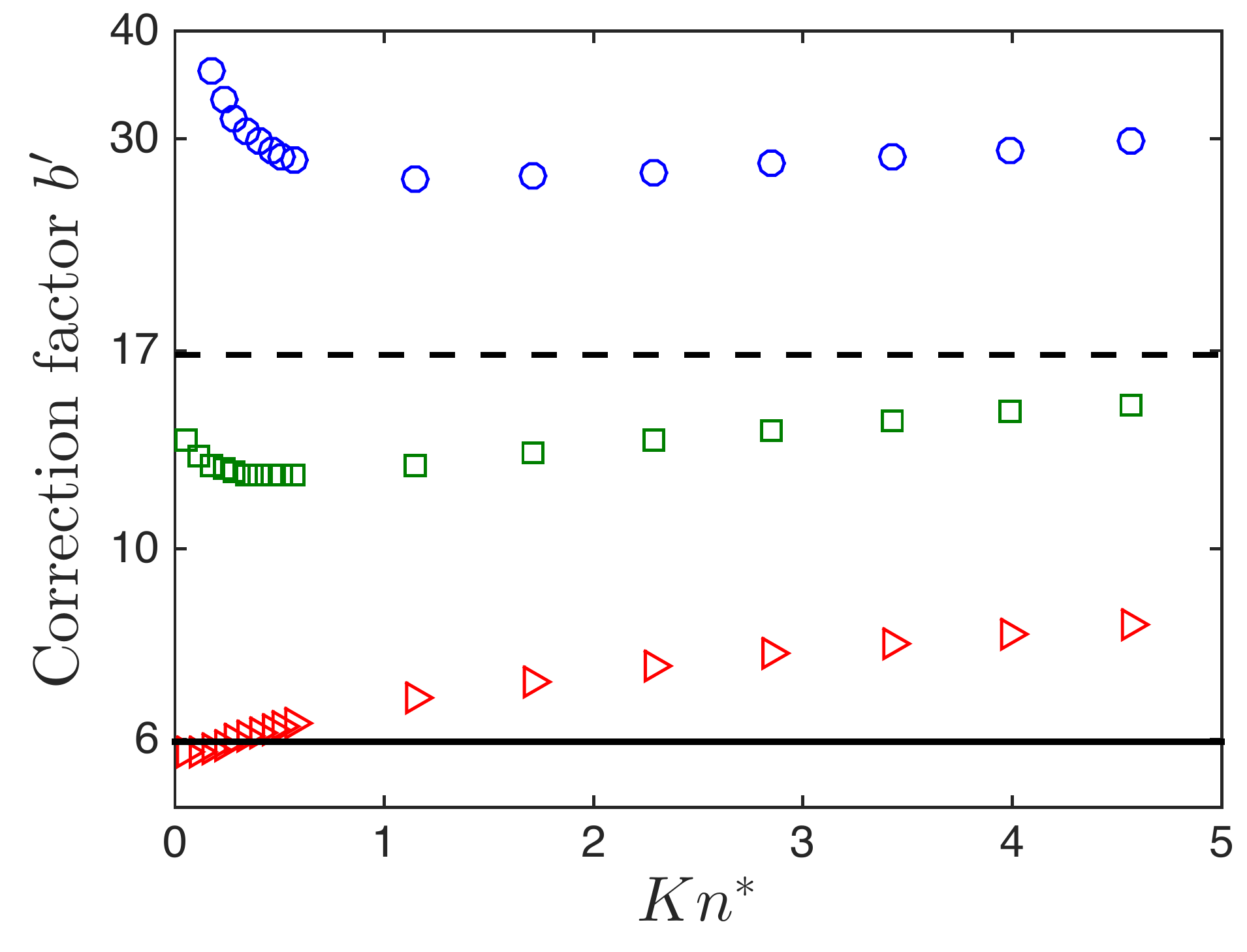}}
	\hskip 0.5cm
	\subfloat[]	{\includegraphics[scale=0.33,viewport=0 200 550 630,clip=true]{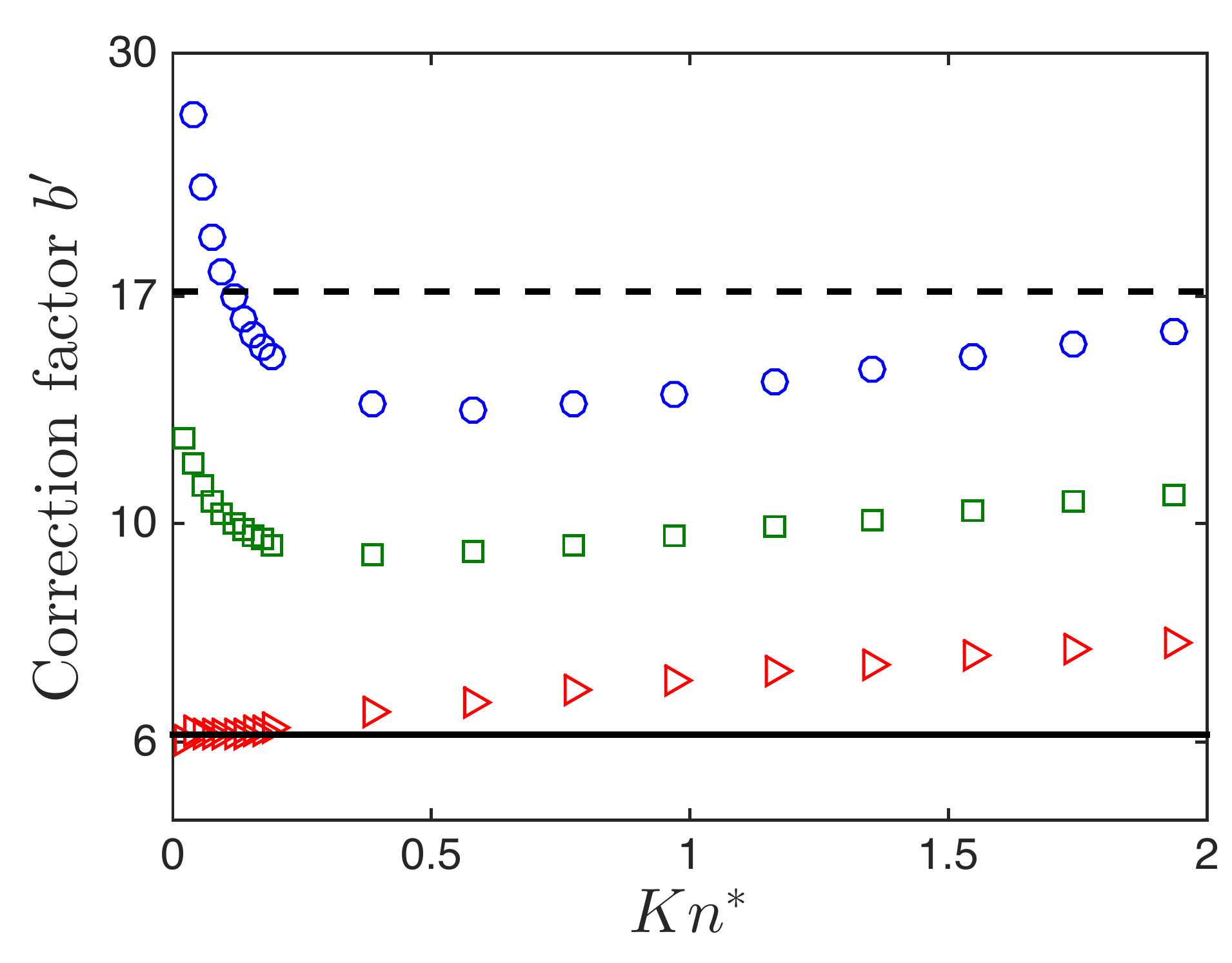}}
	\caption{(First row) Streamlines in  unit computational cells. (Second row) The ratio of the AGP to the intrinsic permeability and (third row) the correction factor, as  functions of $Kn^\ast$. Note that the vertical axis in (e) and (f) is in the logarithmic scale. Triangles, squares, and circles are numerical results of the linearized BGK equation (solved by the discrete velocity method), with the TMAC $\alpha=1.0$, 0.5, and 0.1, respectively. The black lines are the numerical solutions from the NSEs with FVBC, where only up to the first-order terms of $Kn$ are retained.  In the left column, the porosity is $\epsilon=0.4$, $k_\infty=0.001$, and  $Kn^\ast=5.69Kn$, while in the right column, $\epsilon=0.8$, $k_\infty=0.018$, and  $Kn^\ast=1.94Kn$.
	}	
	\label{Square0408}
\end{figure}

Our numerical solutions for the linearized BGK equation, solved by the discrete velocity method, are shown in Fig.~\ref{Square0408}. The slip-corrected permeability (obtained from the numerical solution of the NSEs with FVBC, but only keeping the linear dependence of the AGP with the Knudsen number at small $Kn$) is also shown for comparison. From Fig.~\ref{Square0408}(c,d) we see that, as the numerical results in~\S~\ref{accuracy}, when the TMAC is $\alpha=1$, the slip-corrected permeability agrees well with the numerical results of the linearized BGK equation, when the porosity is $\epsilon=0.4$ and $0.8$. Also, it is interesting to note that, when $Kn^\ast\rightarrow0$, the correction factor $b'$ is about 6 and 6.1 when the porosity is $\epsilon=0.4$ and 0.8, respectively. For Poiseuille flow between two parallel plates, the slip-corrected permeability is exactly
\begin{equation}
\frac{k_a}{k_\infty}=1+6\xi{Kn^\ast},
\end{equation} 
where $\xi=1.15$ for the diffuse boundary condition. This means that the correction factor is $b'=6.9$, which is close to the two values we obtained  for complex porous media, when the gas-surface interaction is diffuse. 

However, when the TMAC is small, the accuracy of the slip-corrected permeability is significantly reduced. Since at $\alpha=0.5$ and 0.1, $\xi$ is respectively $3.23$ and 19.3, the correction factor $b'$ from the NSEs with FVBC is about 17 and 100, respectively. From Fig.~\ref{Square0408}(c,d) we see that the slip-corrected permeability is only accurate when $Kn^\ast<0.1$.  At relative large Knudsen numbers (i.e. $Kn^\ast>0.5$), the accuracy of the slip-corrected permeability is greatly reduced. For instance, when the porosity is 0.8, the slip-corrected permeability overpredicts the AGP by 160\% and 670\%, when $\alpha=0.5$ and 0.1, respectively; when the porosity is 0.4, the slip-corrected permeability overpredicts the AGP by 360\% when $\alpha=0.1$. These two examples clearly show that, when the TMAC deviates largely from one, the NSEs with FVBC should not be used to predict the AGP in porous media.

Now, we focus on the influence of the non-unitary TMAC on the correction factor.  From Fig.~\ref{Square0408}(c,d) we see that, as expected, when $Kn$ is fixed, the AGP increases when TMAC decreases. However, at different $Kn$ the amount of increase is different, so the variation of $b'$ with respect to $Kn$ is quite different among the three TMACs considered. When $\alpha=1$, it is seen that the correction factor $b'$ increases with $Kn$, which clearly contradicts Klinkenberg's experimental results. However, when $\alpha=0.5$ and 0.1, it is found that, when $Kn$ increases, $b'$ first decreases and then increases. The reason that the correction factor increases with the Knudsen number when $Kn$ is large can be easily understood. It is well-known that, when  $Kn\rightarrow\infty$, the dimensionless mass flow rate $G_p$ in~\eqref{apparentP} increases with $Kn$ logarithmically for Poiseuille flow between two parallel plates~\citep{Takata2011} and approaches a constant for Poiseuille flow through the cylindrical tube. Similar behaviors for $G_p$ are observed for all the cases considered in this paper. According to~\eqref{Klinkenberg2}, the equivalent correction factor $b'$ at $Kn^\ast\rightarrow\infty$ can be calculated as follows:
\begin{equation}\label{Klinkenberg3}
b'=\frac{{k_a}/{k_\infty}-1}{Kn^\ast}=\frac{{2KnG_p}/{\sqrt{\pi}k_\infty}}{Kn^\ast}-\frac{1}{Kn^\ast}\equiv{}C-\frac{1}{Kn^\ast},
\end{equation}
where $C$ is proportional to $G_p$. Depending on the structure of the porous medium, $C$ is either a constant or increases with $Kn$ when $Kn\rightarrow\infty$. Therefore, the correction factor increases with $Kn$. Also, since $G_p$ always increases when TMAC decreases, at large $Kn$, the correction factor increases when TMAC decreases, see Fig.~\ref{Square0408}(e,f). 

\begin{figure}
	\centering
	\subfloat[]{\includegraphics[scale=0.6,viewport=30 265 555 575,clip=true]{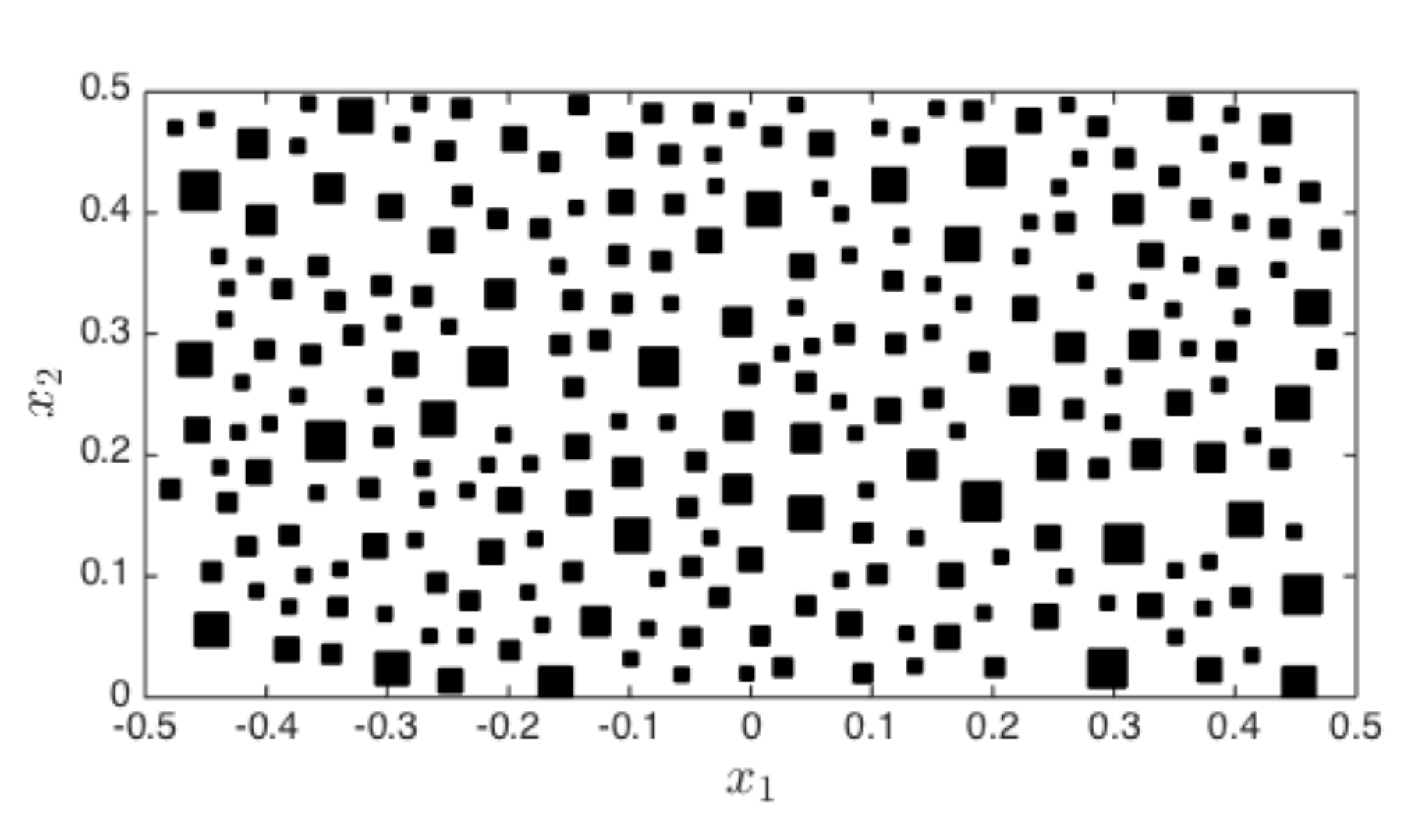}}\\
	
	\subfloat[]	{\includegraphics[scale=0.33,viewport=0 215 550 600,clip=true]{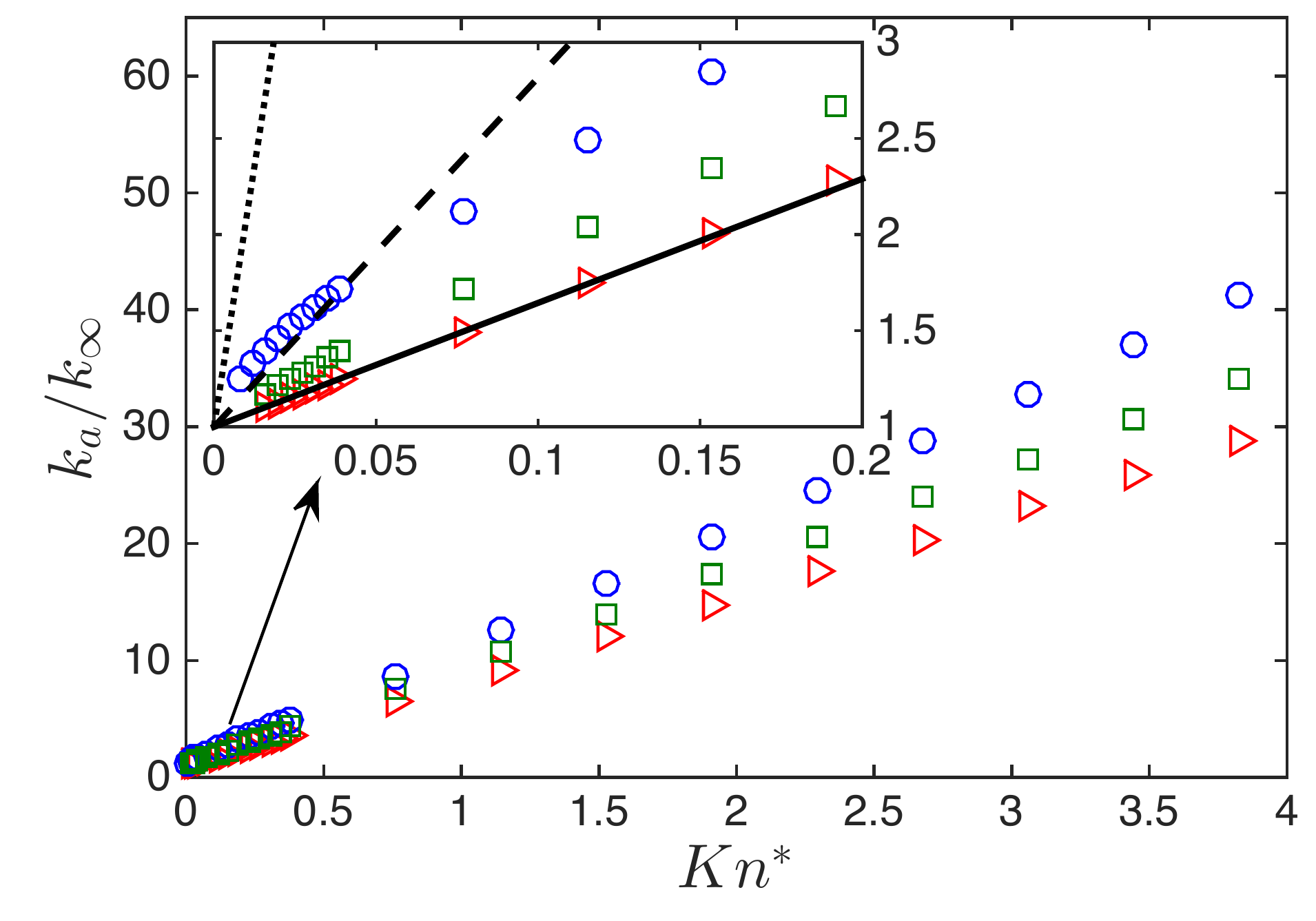}}
	\hskip 0.5cm
	\subfloat[]	{\includegraphics[scale=0.33,viewport=0 215 550 600,clip=true]{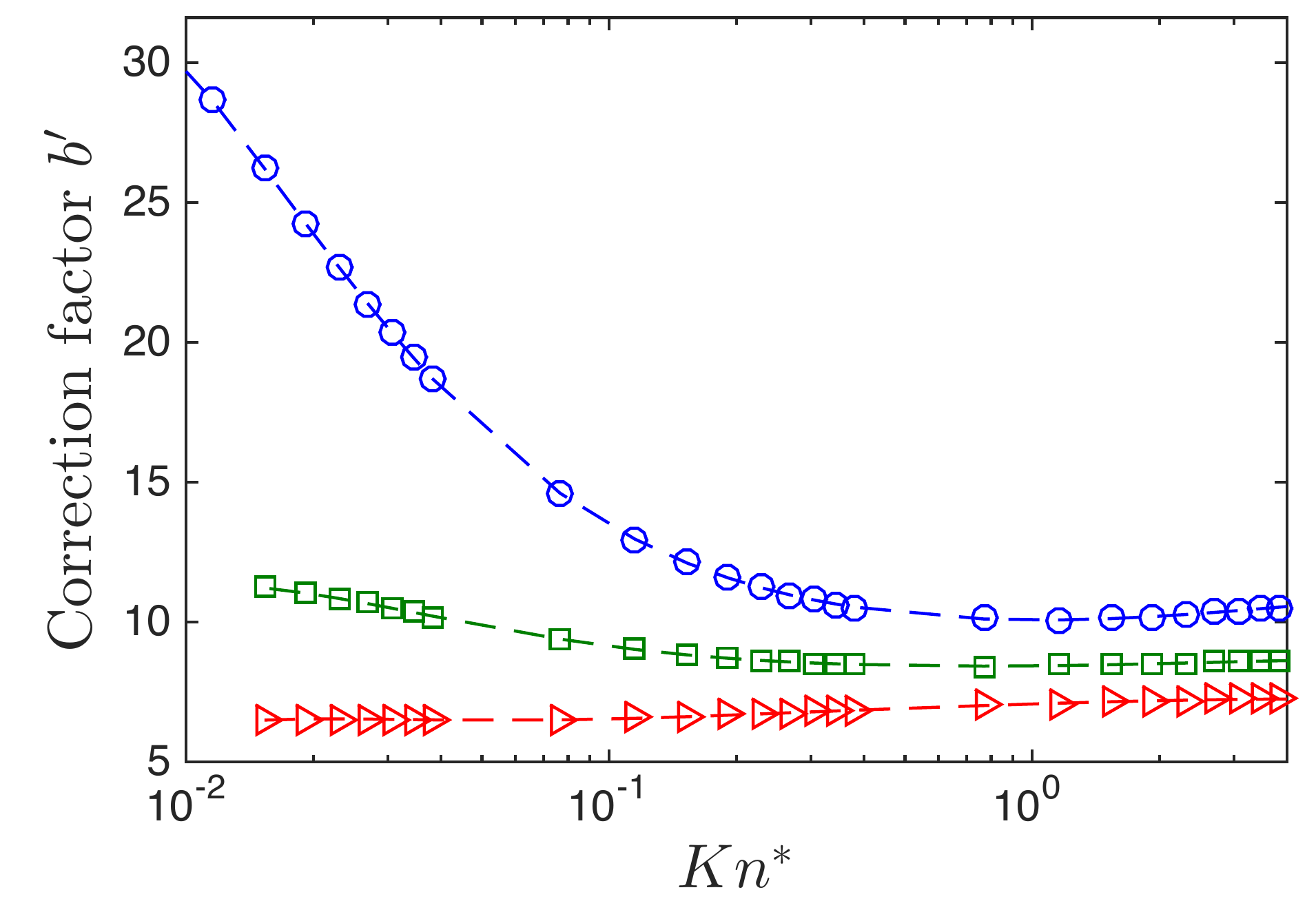}}
	
	\caption{(a) The geometry in a unit computational cell, when the porosity is $\epsilon=0.84$. Solids of random size and position are shown in black. The periodic porous medium is generated by placing the whole computational domain inside the unit rectangular cell ABCD in Fig.~\ref{GEO}. $2000\times1000$ cells are used to discretize the spatial domain. (b) The ratio of the AGP to the intrinsic permeability $k_\infty=4.77\times10^{-5}$ and (c) The correction factor $b'$, as functions of $Kn^\ast$. Triangles, squares, and circles are numerical results of the linearized BGK equation, with the TMAC $\alpha=1.0$, 0.5, and 0.1, respectively. In the inset in (b), the solid, dashed, dotted lines are the slip-corrected permeability for $\alpha=1.0$, 0.5, and 0.1, respectively, where the slopes (correction factor $b'$) are 6.45, 18.2, and 108.5, respectively. Note that $Kn^\ast=38.3Kn$.
	}	
	\label{RandomSquare}
\end{figure}

At small $Kn$, when $\alpha=0.5$ and 0.1, the correction factor decreases when $Kn$ increases. For instance, when the porosity is $\epsilon=0.8$, the correction factor $b'$ with $\alpha=0.5$ drops from 17 when $Kn^\ast\rightarrow0$ to the minimum value 9 at $Kn^\ast=0.4$, while $b'$ with $\alpha=0.1$ drops from 100 at $Kn^\ast\rightarrow0$ to the minimum value 13 at $Kn^\ast=0.6$. When the porosity is $\epsilon=0.4$, the correction factor $b'$ with $\alpha=0.5$ drops from 17 at $Kn^\ast\rightarrow0$ to the minimum value 12 at $Kn^\ast=0.5$, while $b'$ with $\alpha=0.1$ drops from 100 when $Kn^\ast\rightarrow0$ to the minimum value 27 at $Kn^\ast=1.1$. This behavior, is in qualitatively agreement\footnote{Note that a quantitative comparison between the experiment and our simulation is currently impossible because we do not have the detailed structure data of the porous media.} with observations by~\cite{Klinkenberg1941} and~\cite{HeriotWatt2016}.

Our findings also apply to complicated porous media. For example, we have investigated the rarefied gas flow through a porous medium consisting of squares of random size and position in Fig.~\ref{RandomSquare}(a). A similar trend in the variation of the correction factor with respect to the Knudsen number is observed for non-unitary TMACs, see Fig.~\ref{RandomSquare}(c). Also, from the inset of Fig.~\ref{RandomSquare}(b) we see that even the slip-corrected permeability, obtained from the NSEs with FVBC, is only accurate when $k_a/k_\infty\lesssim1.5$ when $\alpha=0.5$ and $0.1$.

So far, we have identified two key factors that could lead to the observation that the correction factor $b$ in~\eqref{Klinkenberg} decreases when $Kn$ (the reciprocal mean gas pressure) increases. The first factor is that the TMAC should be less than 1, and the second one is that flow streamlines should be tortuous. The two factors should be combined together to explain Klinkenberg's observations, as we have shown that the variation of TMAC in the straight cylindrical tube (see Fig.~\ref{ApparentTube}) and the variation of porous media but with TMAC being one (see Figs.~\ref{Apparent_Permeability} and~\ref{RandomDisc}) cannot yield the result that the correction factor decreases when $Kn$ increases.

The reason that Klinkenberg did not observe the increase of the correction factor with the Knudsen number is probably because $Kn$ is not large enough in his experiments, as from Tables 5, 6, and 7 in~\cite{Klinkenberg1941} the maximum value of $k_a/\kappa_\infty$ is around 30. In this region of $k_a/\kappa_\infty$, from Fig.~\ref{Square0408}(c,e) and Fig.~\ref{RandomSquare}(b,c) we see that, roughly, the correction factor $b'$ decreases when $Kn^\ast$ increases.

\section{Conclusions}\label{Conclusions}

The main purpose of this paper is devoted to understanding the Klinkenberg's famous experiment on the apparent gas permeability of the porous media, i.e. to understand why in his experiments the correction factor decreases when the Knudsen number increases. By solving the linearized Bhatnagar-Gross-Krook equation in simplified but not simple porous media via the discrete velocity method, for the first time in the last 76 years, we have pinpointed two key ingredients that lead to his discovery: the tortuous flow path and the coexistence of the diffuse and specular scatterings of the gas molecules when impinging on solid surfaces. Since in experiments the flow paths are always tortuous, Klinkenberg's and others experimental results suggest that the tangential momentum accommodation coefficient for the gas-surface interaction is less than one.

We have also found that Klinkenberg's results can only be observed when the ratio between the apparent and intrinsic permeabilities is $\lesssim30$; at large ratios the correction factor increases with the Knudsen number.

Our numerical solutions not only shed new light on the complex flow physics in porous media, but also could serve as benchmarking cases to check the accuracy of various macroscopic models and/or numerical methods for modeling/simulation of rarefied gas flows in  complex geometries. Specifically, through theoretical and numerical analysis, we have shown that the Navier-Stokes equations with the first-order velocity-slip boundary condition can only predict the apparent gas permeability of porous media to the first-order accuracy of the Knudsen number. Although the slip-corrected expression, which only retains the linear dependence of the permeability and Knudsen number, works fine for the diffuse gas-surface boundary condition in a wide range of gas rarefaction, its accuracy is significantly reduced when the tangential momentum accommodation coefficient $\alpha$ in the diffuse-specular boundary condition deviates from one. For example, when $\alpha=0.5$ and 0.1, the slip-corrected expression is only accurate when the ratio between the apparent and intrinsic permeabilities is $\lesssim1.5$. Thus, the slip-corrected expression is of no practical use. These issues must be properly taken into account since the Navier-Stokes equations are widely misused in rarefied gas flows to predict the apparent gas permeability.

Our work also implies that, all the currently widely used empirical solutions, such as~\eqref{Beskok} derived from straight cylindrical tube with the diffuse boundary condition, should be reformulated to incorporate the tortuosity and diffuse-specular scattering, to predict the unconventional gas production by feeding the apparent gas permeability into upscaling models. This forms our future research directions. Also, it should be noted that all our conclusions are based on the simulation in two-dimensional geometry; whether the results hold or not for three-dimensional porous media requires future systematical investigation.

\section*{Acknowledgements}
LW acknowledges the support of an Early Career Researcher International Exchange Award from the Glasgow Research Partnership in Engineering (allowing him to visit the Hong Kong University of Science and Technology for one month) and the joint project from the Royal Society of Edinburgh and National Natural Science Foundation of China. LG thanks Lianhua Zhu for helpful discussions on  the first-order velocity-slip boundary condition in OpenFoam. This work is also partly supported by the Engineering and Physical Sciences Research Council in the UK under grant EP/M021475/1.


\appendix

\section{Wall boundary conditions for high-order moments}\label{Wall_hob}

The wall boundary conditions for higher-order moments in the regularized 20-moment equations are given as follows:
\begin{equation*}
\begin{split}
\sigma _{\tau \tau} = & - \frac{{2 - \alpha }}{\alpha }\sqrt {\frac{{\pi RT}}{2}} \left( {\frac{{5{m_{n\tau \tau }} + 2{q_n}}}{{5RT}}} \right) + {p_\alpha }\left( {\hat u_\tau ^2 + {{\hat T}_w} - 1} \right) - \frac{{{R_{\tau \tau }} + {R_{nn}}}}{{14RT}} \\
&- \frac{\Delta }{{30RT}} - \frac{{{\phi _{nn\tau \tau }}}}{{2RT}},\\
\sigma _{nn} = & - \frac{{2 - \alpha }}{\alpha }\sqrt {\frac{{\pi RT}}{2}} \left( {\frac{{5{m_{nnn}} + 6{q_n}}}{{10RT}}} \right) + {p_\alpha }\left( {{{\hat T}_w} - 1} \right) - \frac{{{R_{nn}}}}{{7RT}} - \frac{\Delta }{{30RT}} - \frac{{{\phi _{nnnn}}}}{{6RT}},\\
q_\tau = & - \frac{5}{{18}}\frac{{2 - \alpha }}{\alpha }\sqrt {\frac{{\pi RT}}{2}} \left( {7{\sigma _{n\tau }} + \frac{{{R_{n\tau }}}}{{RT}}} \right) - \frac{{5{{\hat u}_\tau }{p_\alpha }\sqrt {RT} \left( {\hat u_\tau ^2 + 6{{\hat T}_w}} \right)}}{{18}} - \frac{{10{m_{nn\tau }}}}{9},\\
m_{\tau \tau \tau } =&  - \frac{{2 - \alpha }}{\alpha }\sqrt {\frac{{\pi RT}}{2}} \left( {3{\sigma _{n\tau }} + \frac{{3{R_{n\tau }}}}{{7RT}} + \frac{{{\phi _{n\tau \tau \tau }}}}{{RT}}} \right) - {p_\alpha }{{\hat u}_\tau }\sqrt {RT} \left( {\hat u_\tau ^2 + 3{{\hat T}_w}} \right) \\
&- \frac{{3{m_{nn\tau }}}}{2} - \frac{{9{q_\tau }}}{5},\\
m_{nn\tau }= & - \frac{{2 - \alpha }}{\alpha }\sqrt {\frac{{\pi RT}}{2}} \left( {{\sigma _{\tau n}} + \frac{{{R_{n\tau }}}}{{7RT}} + \frac{{{\phi _{nnn\tau }}}}{{3RT}}} \right) - \frac{2}{5}{q_\tau } - \frac{{2{{\hat T}_w}{{\hat u}_\tau }{p_\alpha }\sqrt {RT} }}{3}.
\end{split}
\end{equation*}
where ${\hat u_\tau } = u_\tau/\sqrt{RT}$ and ${\hat T_w} = T_w/T$.

\bibliographystyle{jfm}
\bibliography{Bib2}

\begin{thebibliography}{33}
\expandafter\ifx\csname natexlab\endcsname\relax\def\natexlab#1{#1}\fi
\def\au#1{#1} \def\ed#1{#1} \def\yr#1{#1}\def\at#1{#1}\def\jt#1{\textit{#1}}
  \def\bt#1{#1}\def\bvol#1{\textbf{#1}} \def\vol#1{#1} \def\pg#1{#1}
  \def\publ#1{#1}\def\arxiv#1{#1}\def\org#1{#1}\def\st#1{\textit{#1}}

\bibitem[Beskok \& Karniadakis(1999)]{Beskok1999}
{\sc \au{Beskok, A.} \& \au{Karniadakis, G.~E.}} \yr{1999}  \at{A model for
  flows in channels, pipes, and ducts at micro and nano scales}.
  \jt{Microscale Thermophys Eng.}  \bvol{3},  \pg{43--77}.

\bibitem[Bhatnagar {\em et~al.\/}(1954)Bhatnagar, Gross \&
  Krook]{Bhatnagar1954}
{\sc \au{Bhatnagar, P.~L.}, \au{Gross, E.~P.} \& \au{Krook, M.}} \yr{1954}
  \at{A model for collision processes in gases. {I}. {S}mall amplitude
  processes in charged and neutral one-component systems}.  \jt{Phys. Rev.}
  \bvol{94},  \pg{511--525}.

\bibitem[Borner {\em et~al.\/}(2017)Borner, Panerai \& Mansour]{Borner2017}
{\sc \au{Borner, A.}, \au{Panerai, F.} \& \au{Mansour, N.~N.}} \yr{2017}
  \at{{High temperature permeability of fibrous materials using direct
  simulation Monte Carlo}}.  \jt{Int. J. Heat Mass Transfer}  \bvol{106},
  \pg{1318--1326}.

\bibitem[Chai {\em et~al.\/}(2011)Chai, Lu, Shi \& Guo]{Chai2011}
{\sc \au{Chai, Z.}, \au{Lu, J.}, \au{Shi, B.} \& \au{Guo, Z.}} \yr{2011}
  \at{Gas slippage effect on the permeability of circular cylinders in a square
  array}.  \jt{Int. J. Heat Mass Transfer}  \bvol{54},  \pg{3009--3014}.

\bibitem[Chapman \& Cowling(1970)]{CE}
{\sc \au{Chapman, S.} \& \au{Cowling, T.~G.}} \yr{1970} {\em {The Mathematical
  Theory of Non-uniform Gases}\/}.  \publ{Cambridge University Press}.

\bibitem[Civan(2010)]{Civan2010}
{\sc \au{Civan, F.}} \yr{2010}  \at{Effective correlation of apparent gas
  permeability in tight porous media}.  \jt{Transp. Porous. Med.}  \bvol{82},
  \pg{375--384}.

\bibitem[Darabi {\em et~al.\/}(2012)Darabi, Ettehad, Javadpour \&
  Sepehrnoori]{Javadpour2012}
{\sc \au{Darabi, H.}, \au{Ettehad, A.}, \au{Javadpour, F.} \& \au{Sepehrnoori,
  K.}} \yr{2012}  \at{Gas flow in ultra-tight shale strata}.  \jt{J. Fluid
  Mech.}  \bvol{710},  \pg{641--658}.

\bibitem[Garcia-Colin {\em et~al.\/}(2008)Garcia-Colin, Velasco \&
  Uribe]{Colin2008}
{\sc \au{Garcia-Colin, L.~S.}, \au{Velasco, R.~M.} \& \au{Uribe, F.~J.}}
  \yr{2008}  \at{Beyond the {N}avier-{S}tokes equations: {Burnett}
  hydrodynamics}.  \jt{Phys. Rep.}  \bvol{465},  \pg{149--189}.

\bibitem[Gibelli(2011)]{Gibelli_2ndSLIP2011}
{\sc \au{Gibelli, L.}} \yr{2011} A second-order slip model for arbitrary
  accomodation at the wall.  \bt{In {\em 3rd Micro and Nano Flows\/}}.
  \publ{Brunel University}.

\bibitem[Grad(1949)]{Grad1949}
{\sc \au{Grad, H.}} \yr{1949}  \at{On the kinetic theory of rarefied gases}.
  \jt{Comm. Pure Appl. Math.}  \bvol{2},  \pg{331--407}.

\bibitem[Gu \& Emerson(2009)]{Gu2009}
{\sc \au{Gu, X.~J.} \& \au{Emerson, D.~R.}} \yr{2009}  \at{A high-order moment
  approach for capturing non-equilibrium phenomena in the transition regime}.
  \jt{J. Fluid Mech.}  \bvol{636},  \pg{177--216}.

\bibitem[Gu {\em et~al.\/}(2010)Gu, Emerson \& Tang]{GuPRE2010}
{\sc \au{Gu, X.~J.}, \au{Emerson, D.~R.} \& \au{Tang, G.~H.}} \yr{2010}
  \at{{Analysis of the slip coefficient and defect velocity in the Knudsen
  layer of a rarefied gas using the linearized moment equations}}.  \jt{Phys.
  Rev. E}  \bvol{81},  \pg{016313}.

\bibitem[Hadjiconstantinou(2003)]{Hadjiconstantinou2003slip}
{\sc \au{Hadjiconstantinou, N.~G.}} \yr{2003}  \at{{Comment on Cercignani's
  second-order slip coefficient}}.  \jt{Phys. Fluids}  \bvol{15},
  \pg{2352--2354}.

\bibitem[Huang {\em et~al.\/}(2013)Huang, Xu \& Yu]{Huang2013}
{\sc \au{Huang, J.~C.}, \au{Xu, K.} \& \au{Yu, P.~B.}} \yr{2013}  \at{A unified
  gas-kinetic scheme for continuum and rarefied flows {III}: Microflow
  simulations}.  \jt{Commun. Comput. Phys.}  \bvol{14},  \pg{1147--1173}.

\bibitem[Karniadakis {\em et~al.\/}(2005)Karniadakis, Beskok \&
  Aluru]{Beskok_book}
{\sc \au{Karniadakis, G.}, \au{Beskok, A.} \& \au{Aluru, N.~R.}} \yr{2005} {\em
  {Microflows and Manoflows: Fundamentals and Simulations}\/}.  \publ{Springer,
  New York}.

\bibitem[Klinkenberg(1941)]{Klinkenberg1941}
{\sc \au{Klinkenberg, L.~J.}} \yr{1941}  \at{The permeability of porous media
  to liquids and gases}.  \jt{American Petroleum Institute}  \pg{pp.
  API--41--200}.

\bibitem[Lasseux {\em et~al.\/}(2016)Lasseux, {Valdes Parada} \&
  Porter]{Didier2016}
{\sc \au{Lasseux, D.}, \au{{Valdes Parada}, F.~J.} \& \au{Porter, M.~L.}}
  \yr{2016}  \at{An improved macroscale model for gas slip flows in porous
  media}.  \jt{J. Fluid Mech.}  \bvol{805},  \pg{118--146}.

\bibitem[Loyalka {\em et~al.\/}(1975)Loyalka, Petrellis \&
  Storvick]{Loyalka1975}
{\sc \au{Loyalka, S.~K.}, \au{Petrellis, N.} \& \au{Storvick, T.~S.}} \yr{1975}
   \at{{Some numerical results for the BGK model - Thermal creep and viscous
  slip problems with arbitrary accomodation at the surface}}.  \jt{Phys.
  Fluids}  \bvol{18},  \pg{1094--1099}.

\bibitem[Lunati \& Lee(2014)]{Lunati2014}
{\sc \au{Lunati, I.} \& \au{Lee, S.~H.}} \yr{2014}  \at{A dual-tube model for
  gas dynamics in fractured nanoporous shale formations}.  \jt{J. Fluid Mech.}
  \bvol{757},  \pg{943--971}.

\bibitem[Maxwell(1879)]{Maxwell1879}
{\sc \au{Maxwell, J.~C.}} \yr{1879}  \at{On stresses in rarefied gases arising
  from inequalities of temperature}.  \jt{Philosophical Transactions of the
  Royal Society Part 1}  \bvol{170},  \pg{231--256}.

\bibitem[Moghaddam \& Jamiolahmady(2016)]{HeriotWatt2016}
{\sc \au{Moghaddam, R.~N.} \& \au{Jamiolahmady, M.}} \yr{2016}  \at{Slip flow
  in porous media}.  \jt{Fuel}  \bvol{173},  \pg{298--310}.

\bibitem[Pan {\em et~al.\/}(2006)Pan, Luo \& Miller]{Pan2006}
{\sc \au{Pan, C.}, \au{Luo, L.~S.} \& \au{Miller, C.~T.}} \yr{2006}  \at{{An
  evaluation of lattice Boltzmann schemes for porous medium flow simulation}}.
  \jt{Comput. \& Fluids}  \bvol{35},  \pg{898--909}.

\bibitem[Sharipov \& Graur(2012)]{GraurVacuum2012}
{\sc \au{Sharipov, F.} \& \au{Graur, I.~A.}} \yr{2012}  \at{Rarefied gas flow
  through a zigzag channel}.  \jt{Vacuum}  \bvol{86},  \pg{1778--1782}.

\bibitem[Struchtrup(2005)]{henning}
{\sc \au{Struchtrup, H.}} \yr{2005} {\em Macroscopic Transport Equations for
  Rarefied Gas Fows: Approximation Methods in Kinetic Theory\/}.
  \publ{Heidelberg, Germany: Springer}.

\bibitem[Struchtrup \& Torrihon(2003)]{Struchtrup2003}
{\sc \au{Struchtrup, H.} \& \au{Torrihon, M.}} \yr{2003}  \at{Regularization of
  {Grad's} 13 moment equations: derivation and linear analysis}.  \jt{Phys.
  Fluids}  \bvol{15},  \pg{2668--2680}.

\bibitem[Taheri {\em et~al.\/}(2009)Taheri, Rana, Torrilhon \&
  Struchtrup]{Taheri2009}
{\sc \au{Taheri, P.}, \au{Rana, A.~S.}, \au{Torrilhon, M.} \& \au{Struchtrup,
  H.}} \yr{2009}  \at{Macroscopic description of steady and unsteady
  rarefaction effects in boundary value problems of gas dynamics}.  \jt{Cont.
  Mech. Theromodyn.}  \bvol{21},  \pg{423--443}.

\bibitem[Takata \& Funagane(2011)]{Takata2011}
{\sc \au{Takata, S.} \& \au{Funagane, H.}} \yr{2011}  \at{Poiseuille and
  thermal transpiration flows of a highly rarefied gas: over-concentration in
  the velocity distribution function}.  \jt{J. Fluid Mech.}  \bvol{669},
  \pg{242--259}.

\bibitem[Tang {\em et~al.\/}(2013)Tang, Zhai, Tao, Gu \& Emerson]{Tang2013CCP}
{\sc \au{Tang, G.~H.}, \au{Zhai, G.~X.}, \au{Tao, W.~Q.}, \au{Gu, X.~J.} \&
  \au{Emerson, D.~R.}} \yr{2013}  \at{Extended thermodynamic approach for
  non-equilibrium gas flow}.  \jt{Commun. Comput. Phys.}  \bvol{13},
  \pg{1330--1356}.

\bibitem[Torrilhon(2016)]{TorrihonReview2016}
{\sc \au{Torrilhon, M.}} \yr{2016}  \at{{Modeling nonequilibrium gas flow based
  on moment equations}}.  \jt{Annn. Rev. Fluid Mech.}  \bvol{48},
  \pg{429--458}.

\bibitem[Wang {\em et~al.\/}(2014)Wang, Chen, Jha \& Rogers]{Wang2014Shale}
{\sc \au{Wang, Q.}, \au{Chen, X.}, \au{Jha, A.} \& \au{Rogers, H.}} \yr{2014}
  \at{{Natural gas from shale formation - The evolution, evidences and
  challenges of shale gas revolution in United States}}.  \jt{Renew. Sust.
  Energ. Rev.}  \bvol{30},  \pg{1--28}.

\bibitem[Wu {\em et~al.\/}(2014)Wu, Reese \& Zhang]{lei_Jfm}
{\sc \au{Wu, L.}, \au{Reese, J.~M.} \& \au{Zhang, Y.~H.}} \yr{2014}
  \at{{Solving the Boltzmann equation by the fast spectral method: application
  to microflows}}.  \jt{J. Fluid Mech.}  \bvol{746},  \pg{53--84}.

\bibitem[Wu {\em et~al.\/}(2013)Wu, White, Scanlon, Reese \& Zhang]{lei}
{\sc \au{Wu, L.}, \au{White, C.}, \au{Scanlon, T.~J.}, \au{Reese, J.~M.} \&
  \au{Zhang, Y.~H.}} \yr{2013}  \at{{Deterministic numerical solutions of the
  Boltzmann equation using the fast spectral method}}.  \jt{J. Comput. Phys.}
  \bvol{250},  \pg{27--52}.

\bibitem[Zhu {\em et~al.\/}(2016)Zhu, Zhong \& Xu]{zhuyajun2016}
{\sc \au{Zhu, Y.}, \au{Zhong, C.} \& \au{Xu, K.}} \yr{2016}  \at{Implicit
  unified gas-kinetic scheme for steady state solutions in all flow regimes}.
  \jt{J. Comput. Phys.}  \bvol{315},  \pg{16--38}.

\end{thebibliography}

\end{document}